\pdfoutput=1
\documentclass[prl,reprint, epsfig,amssymb,bibliography,amsmath,footinbib]{revtex4-2}

\usepackage{blkarray}

\usepackage{amssymb,amsmath,bm,epsfig,graphicx,subfigure,hyperref}

\usepackage[utf8]{inputenc}
\usepackage[vietnamese, english]{babel}
\usepackage{chngcntr}
\usepackage{xcolor}
\usepackage{braket}
\usepackage{physics}
\usepackage{sidecap}
\usepackage{lipsum}

\begin{document}

\title{Boundary Modes from Periodic Magnetic and Pseudomagnetic Fields in Graphene}
\author{\foreignlanguage{vietnamese}{Võ Tiến Phong}}
\author{E. J. Mele}
\email{mele@physics.upenn.edu}
\affiliation{Department of Physics and Astronomy, University of Pennsylvania, Philadelphia PA 19104}
\date{\today}

\begin{abstract}
Single-layer graphenes subject to periodic lateral strains are artificial crystals that can support boundary spectra with an intrinsic polarity.  These are analyzed by comparing the effects of periodic magnetic fields and  strain-induced pseudomagnetic fields that respectively break and preserve time-reversal symmetry. In the former case, a Chern classification of the superlattice minibands with zero total magnetic flux enforces {\it single} counter-propagating modes traversing each bulk gap on opposite boundaries of a nanoribbon. For the pseudomagnetic field, pairs of counter-propagating modes migrate to the {\it same} boundary where they provide well-developed valley-helical transport channels on a single zigzag edge. We discuss possible schemes for implementing this situation and their experimental signatures.
\end{abstract}

\maketitle

Atomically-thin materials are versatile platforms for creating new artificial crystals defined by imposing laterally-periodic potentials.  A celebrated example is twisted bilayer graphene where a small interlayer twist modulates the  atomic registry on large moir\'e scales $L \sim 10-20 \, {\rm nm}$ and yields structurally-tunable superperiodic solids with flattened minibands \cite{CF18,C18}. When the bandwidth is made much smaller than the energy gaps separating manifolds, the electronic physics projected into the spectrally-isolated minibands can be controlled by various many-body effects \cite{AM20, BD20}. Fractured spectra can be produced even for a monolayer material by other mechanisms that introduce  Bragg reflections on a superlattice, such as by periodic patterning of an electrostatic gate or a perpendicular magnetic field \cite{R20,ST21}. The latter situation is interesting since even if the spatially-averaged flux vanishes, the applied magnetic field breaks time-reversal symmetry ${\cal T}$, and can generate gapped minibands with non-zero Chern numbers \cite{TKNN82,H88}. However, experimentally realizing the anomalous Hall effect via this route is daunting because of the large  magnetic fields required to produce sufficiently large minigaps. Nonetheless, it suggests an alternate approach where instead a periodic strain field in monolayer graphene is coupled to electronic motion and masquerades as a valley-antisymmetric pseudomagnetic field  on a much larger energy scale.  This has been demonstrated experimentally for a sheet of graphene contacted with NbSe$_2$ \cite{MM20}. The properties of nearly flat bands in this system have been probed for possible correlated-electron physics \cite{MM20, MA20, ML20, ML21,GCT21}.

Pseudomagnetic fields induce fascinating electronic phenomena even at the single-particle level associated with the symmetries of the superlattice minigaps \cite{CNG09,LB10, GK10, LB10,NR14,RN17,SS16, MS20,BNG20}.   Unlike a real magnetic field, the pseudofield is ${\cal T}$-symmetric and changes its sign in two momentum-space valleys. Each valley can host nontrivial Chern minibands, but this is exactly compensated in the time-reversed valley. Naively, one might expect this to exclude any interesting structure in the boundary spectra. However, we find that this setup still supports rich edge-state physics in its boundary spectra without topological protection \cite{LG10}.  Robust transport in edge channels {\it on selected boundaries} is generically produced by a competition between bulk and surface symmetries when such an artificial crystal is terminated. Below, we analyze the resulting edge modes, demonstrate how they produce an intrinsic polarity on a generic terminated superlattice, and suggest how these features can be probed in transport and spectroscopic measurements.

\begin{figure}[h!]
    \centering
    \includegraphics[width=3.35in]{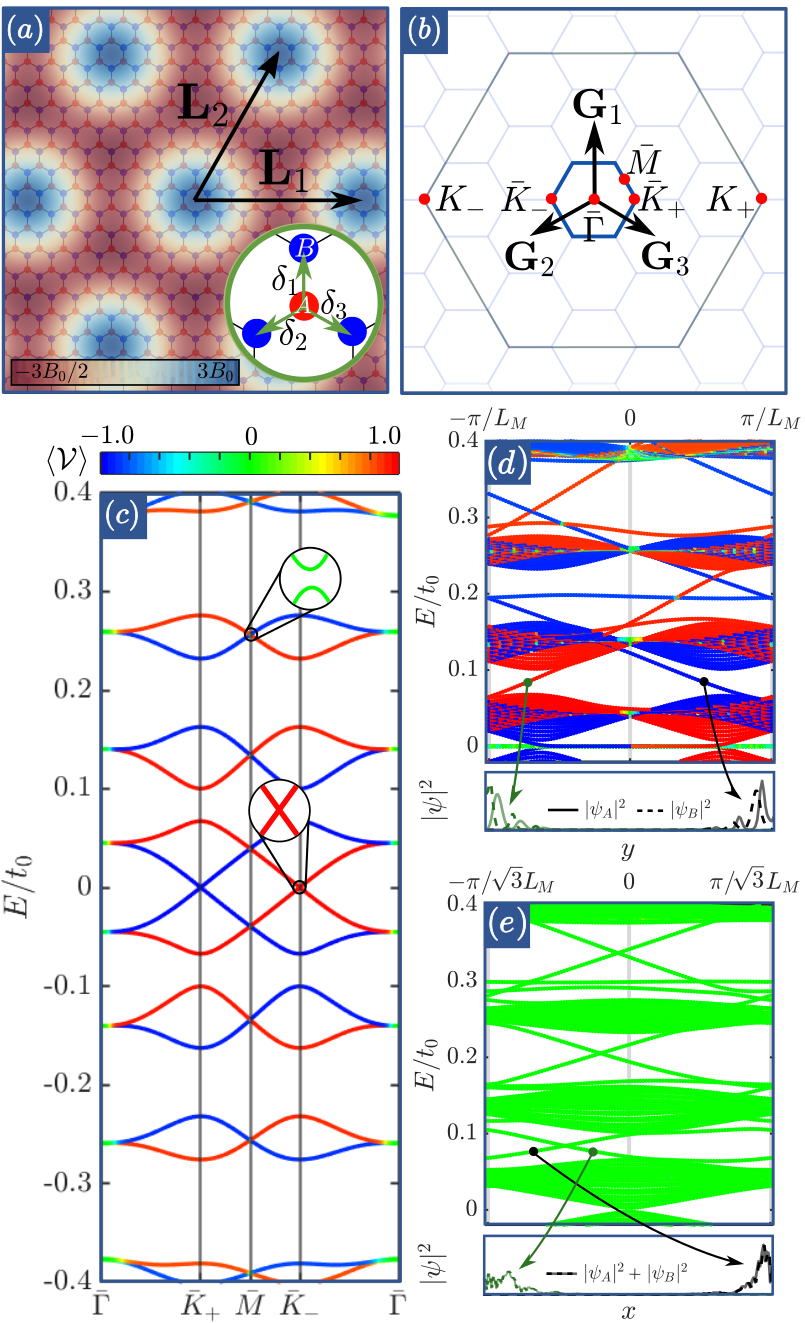}
    \caption{\textbf{Graphene  in a periodically-modulated magnetic field.} (a) Real-space and (b) reciprocal-space representations of graphene in a periodic magnetic field. (c) Band structure for $N = 14$ and $\delta t = 0.3.$ The energy eigenstates are color-coded by their valley expectation value $\langle \mathcal{V}\rangle $ as in Ref. \cite{ML20}. We observe that the bands close to $E=0$ are exceptionally well valley-polarized. Within each valley, there is a doublet band at $E=0,$ and singlet bands elsewhere.  Band structures on a finite zigzag (d) and armchair (e) nanoribbon showing protected topological edge states at every insulating gap shown.  In the zigzag configuration, we see that the edge states are highly valley-polarized. Electron density distributions for some representation edge states are shown below each band structure. Here, each ribbon has $175$ carbon atoms across its width. }
    \label{fig: 1}
\end{figure}

 Before analyzing the band structure of monolayer graphene in a periodic strain field, we begin by considering the closely-related system of \textit{spinless} electrons in a \textit{real} but periodic magnetic field. In this case, the superlattice bands admit a Chern topological classification even without requiring valley projection. Consequently, this allows us to identify robust edge modes that cannot be gapped out by intervalley coupling. We choose a real magnetic field that is spatially varying with a periodic profile and zero total flux, $   \mathbf{B}(\mathbf{r}) = B_0 \hat{z} \sum_{i=1}^3 \cos \left(\mathbf{G}_i \cdot \mathbf{r}  \right) ,$ where $\mathbf{G}_i$ are the superlattice primitive reciprocal lattice vectors specified in Fig. \ref{fig: 1}b. For simplicity, the superlattice translation vectors, $\mathbf{L}_{i} = N \mathbf{a}_{i},$ are chosen to be commensurate with the graphene translation vectors, $\mathbf{a}_i,$ for some integer $N.$

In the presence of a periodic potential, the microscopic Brillouin zone (BZ) folds back to a smaller superlattice Brillouin zone (sBZ). For small $N,$ there are two classes of structures: if $\mod\left( N,3\right) = 0,$ $K_+$ and $K_-$ are mapped to $\Bar{\Gamma}$, and otherwise, $K_+$ and $K_-$ are mapped to different points. For large $N,$ it becomes impractical to distinguish between these two classes of structures because of an emergent valley symmetry that allows us to approximately label states by a valley index. In this limit, we obtain a series of \textit{isolated} flat bands near charge neutrality. Within a single valley, there is one two-dimensional manifold that crosses $E=0,$ and one-dimensional manifolds elsewhere near $E=0$. We use $N = 14$ for efficient numerical calculations throughout this work, though conclusions reached here are applicable to other cases of larger $N$ as well. A representative band structure in this limit for $N=14$ is shown in Fig. \ref{fig: 1}c.  To characterize the band topology of this ${\cal T}$-broken system, we calculate the Chern number associated with every insulating gap close to charge neutrality numerically using a finite difference method \cite{F05, V18}. In our case, since the gaps induced by intervalley hybridization are quite small, a sufficiently fine $\mathbf{k}$-space mesh is necessary in order to obtain convergent results. We find that all the insulating bulk gaps near charge neutrality carry $\mathcal{C} = \text{sign} (B_0E).$

The presence of a nonzero Chern number requires the existence of a dispersive band in the bulk gap at every boundary termination for a macroscopic sample. In an armchair termination, $K_+$ and $K_-$ are mapped onto the same wavector of the ribbon Hamiltonian, so we do not expect edge modes to carry definite valley character. In a zigzag termination,  $K_+$ and $K_-$ are  well separated in the boundary-projected crystal momentum. Consequently, we expect edge modes in this case to be valley-polarized close to charge neutrality. This is confirmed numerically, as shown in Fig. \ref{fig: 1}de. Interestingly, because $\mathcal{C} = \pm 1$ for {\it all} the relevant insulating gaps, there is only {\it one} topological edge mode per boundary in each bulk gap. The boundary states in a zigzag nanoribbon from different valleys belong to opposite edges. This situation is very similar to the edge modes associated with the famous $n=0$ Landau level of a Dirac Hamiltonian  where each valley contributes an edge state at a different edge \cite{BF06,ALL06}. However, the present system with a periodic magnetic field is distinct from the Landau-level scenario because of the absence of prominent insulating gaps with higher Chern numbers. In the Landau-level problem, each band carries a nonzero Chern number, and  the number of edge modes thus increases in the higher gaps. The valley-polarized edge states associated with these higher-Chern-number Landau-level gaps occur on both boundaries of a zigzag nanoribbon.


Using the insights from studying a periodic magnetic field, we now examine graphene under a periodic strain field. Under such a field, the bond strengths between carbon atoms are periodically modulated in space. Again, in the limit $N \gg 1,$ we expect \textit{emergent} superlattice-scale symmetries to dominate the low-energy physics. For a sufficiently smooth superlattice potential, it is useful to approximate the physics near charge neutrality by considering \textit{independent} valley Dirac fermions at $K_+$ and $K_-.$ Throughout this work, we neglect the scalar contribution of the strain field by assuming that only nearest-neighbor hoppings are significant and  the substrate is chemically and electronically inert \cite{BWG18}. In this theory, the strain field enters the Hamiltonian as a spatially-dependent pseudo-gauge field. Inspired by a recent experiment \cite{MM20}, we choose the pseudomagnetic field to be the same as before in one of the valleys, $\mathbf{B}_\nu(\mathbf{r}) = \nu B_0 \hat{z} \sum_{i=1}^3
 \cos (\mathbf{G}_i \cdot \mathbf{r}),$ where $\nu = \pm$ denotes valleys. The valley-projected, long-wavelength Hamiltonian is
\begin{equation}
\label{eq: hamil}
\begin{split}
\hat{\mathcal{H}}_\nu &= \hbar v_F\int d^2 \mathbf{r} \hat{\psi}^\dagger(\mathbf{r})  \left( -i \nabla_\mathbf{r} + \frac{e}{\hbar} \mathbf{A}_\nu(\mathbf{r}) \right) \cdot \left( \nu \sigma_x, \sigma_y \right) \hat{\psi}(\mathbf{r}), 
 \end{split}
\end{equation}
where $\hat{\psi}^\dagger(\mathbf{r}) = \left( \hat{\psi}_A^\dagger(\mathbf{r}),\hat{\psi}_B^\dagger(\mathbf{r}) \right) $ are the sublattice creation operators and $\sigma_j$ act on the sublattice degrees of freedom. In our convention, the charge $q$ is $q=-e$. Under a gauge transformation $\mathbf{A}_\nu(\mathbf{r}) \mapsto \mathbf{A}_\nu(\mathbf{r}) - \nabla f(\mathbf{r}),$ the wavefunctions simply acquire a local $U(1)$ phase $\psi(\mathbf{r}) \mapsto \exp\left( i e f(\mathbf{r})/\hbar \right) \psi(\mathbf{r}).$  This valley-projected Hamiltonian is invariant under the magnetic point symmetry group $3 \underline{m},$ with three classes $\lbrace E, C_{3z}, {\cal T} M_x \rbrace$ \cite{SM, T03,DD07}. When both valleys are considered, then the full Hamiltonian respects $C_{3z},$ $M_x,$ and ${\cal T}$. Importantly, $C_{2z}$ is broken in this configuration.

\begin{figure}[h!]
    \centering
    \includegraphics[width=3.3in]{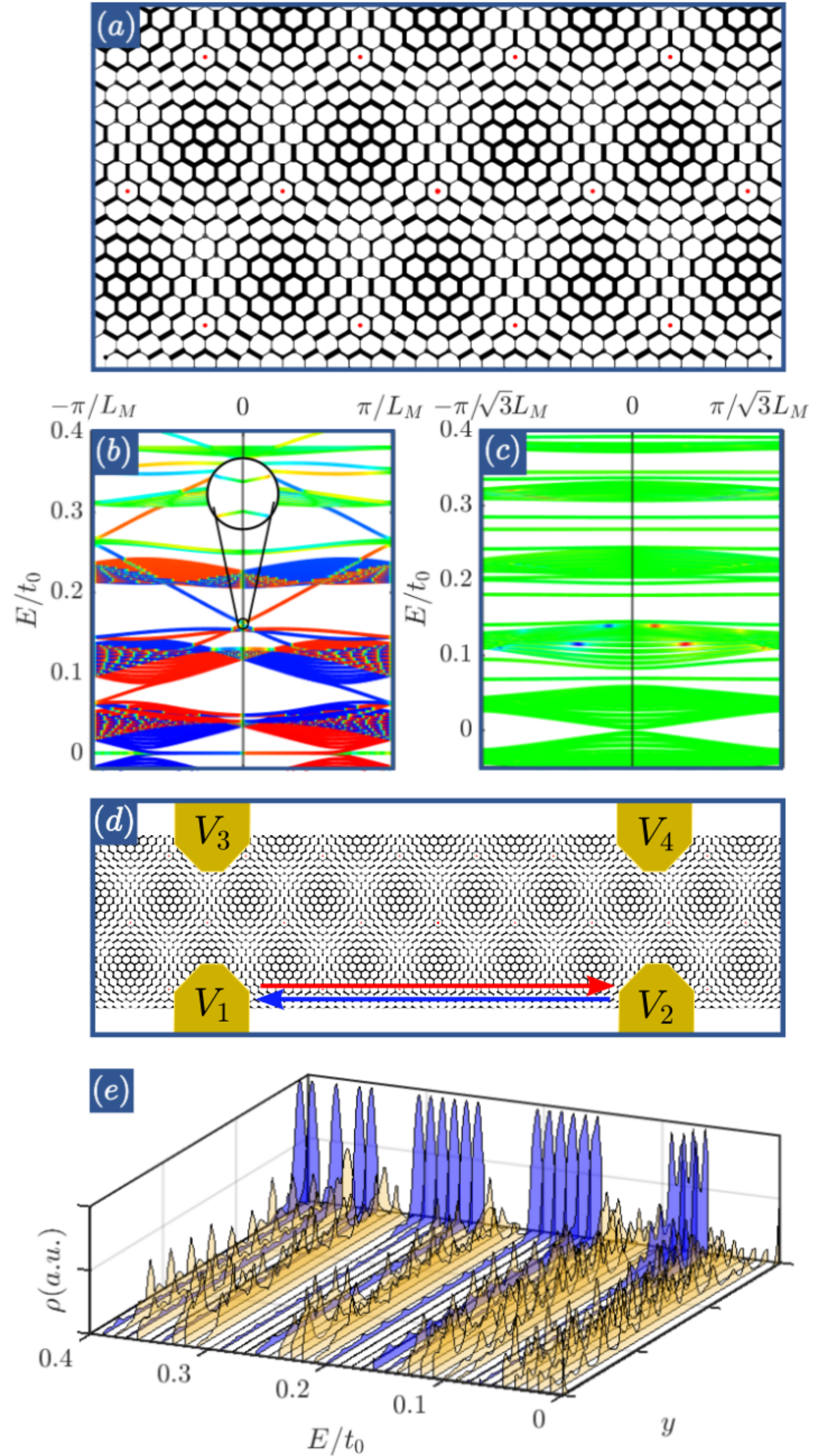}
    \caption{\textbf{Graphene in the presence of a periodic strain.} (a) Lattice realization of the gauge field defined in the text with the bond strength indicated by the thickness of the bonds. In this gauge, $C_{3z}$ and $M_x$ are evidently preserved. Band structures on a finite zigzag (b) nanoribbon showing  edge states at every insulating gap. These edge states sometimes meet at $\mathcal{T}$-invariant momenta and form small avoided crossings. There are no corresponding edge modes in the armchair configuration (c).  The bands are colored according to their valley character, as detailed in Fig. \ref{fig: 1}. Here, $N=14,$ $\delta t = 0.3,$ and each ribbon has $175$ carbon atoms across its width. (d) A four-terminal setup to measure the existence of the one-sided boundary states when the chemical potential resides within one of the bulk gaps. When a potential difference is applied across $V_1$ and $V_2,$ a current flows across the two terminals. On the other hand, when a potential difference is applied across $V_3$ and $V_4,$ no current is detected. (e) Local density of states as a function of energy $E$ and position $y.$ The midgap states shown in blue are localized to only one edge. The bulk states shown in yellow are extended throughout the sample. For this simulation, the ribbon has $150$ carbon atoms across its width.}
    \label{fig: 3}
\end{figure}

Just as in the case of a real magnetic field, the presence of a periodic potential produced by strain fractures the energy spectrum into a series of narrow bands. In the limit of exact valley symmetry, we learn from the case of a real magnetic field that each valley contributes a single edge mode at the insulating gaps away from $E=0$ on a particular boundary of a zigzag nanoribbon. If valley mixing is prohibited, these edge modes must also exist in graphene under a periodic strain. However, due to ${\cal T}$ symmetry, the two edge modes coming from both valleys now must populate the {\it same} zigzag boundary. In other words, the sign change of the pseudomagnetic field between valleys is realized by having the edge states reside on only {\it one side} of a zigzag nanoribbon. Of course, valley symmetry is not strictly exact in a microscopic theory. However, one can effectively suppress intervalley scattering by increasing the superlattice period and operating in the ballistic limit, both of which are achievable in graphene-based platforms. For these reasons, we expect these edge modes to be robust and allow their experimental detection and manipulation although they are not topologically protected.

To confirm the existence of these boundary states numerically, we construct a tight-binding representation of Hamiltonian \eqref{eq: hamil} which respects all of the aforementioned symmetries. Practically, the exact strain field is seldom  determined in an experiment. So we use a desired pseudomagnetic field as the starting point and find a suitable corresponding tight-binding parameterization. To do so, we employ the following approximation $ev_F \left[A_x(\mathbf{R}_j)+ i A_y(\mathbf{R}_j) \right] \approx -\sum_{i=1}^3 \delta t_i\left(\mathbf{R}_j\right) e^{-i \mathbf{K}_+ \cdot \boldsymbol{\delta}_i},$ where $\delta t_i\left(\mathbf{R}_j \right) $ is the bond strength modulation along the $\boldsymbol{\delta}_i$ direction as shown in Fig. \ref{fig: 1}a \cite{KN08,PCN09,MMP13}. Among the many possible choices for $\delta t_i\left(\mathbf{R}_j \right),$ we adopt the following gauge that respects all the spatial symmetries of the continuum model , $    \delta t_i(\mathbf{R}_j) = t_0 \delta t \sin \left(\mathbf{G}_i \cdot \mathbf{R}_j  \right),$ where $t_0\delta t = \sqrt{3}v_F e B_0 L/4\pi$ \cite{ML20, ML21}. A realization in this gauge is shown in Fig. \ref{fig: 3}a, where it is immediately clear that $C_{3z}$ and $M_x$ are preserved. Importantly, $C_{2z}$ is broken explicitly.  If it were preserved, the accumulation of edge modes on only one edge would be impossible when the edge termination is $C_{2z}$-symmetric. This would render the existence of the boundary-selective edge modes dependent on the precise width of a zigzag nanoribbon.

 At low energies, we have confirmed numerically that the bands calculated from the continuum model qualitatively match those produced by the tight-binding model. We then diagonalize the Hamiltonian on finite ribbons with zigzag and armchair terminations. Indeed, in the zigzag configuration where intervalley scattering is suppressed, we find edge states residing in the insulating gaps away from charge neutrality, as shown in Fig. \ref{fig: 3}b. The branches from opposite valleys have small avoided crossings at ${\cal T}$-invariant momenta.  These one-sided boundary states can be detected in a transport measurement using a four-terminal setup as shown in Fig. \ref{fig: 3}d: when the chemical potential is in a bulk gap, one of the two edges acts as an insulator while the other is a conductor.  Because of this,  while the phase with a real magnetic field is characterized by quantized Hall conductivity, this time-reversal symmetric phase necessarily has zero Hall conductivity, but carries quantized longitudinal conductance on the active edge in the clean limit. Another possible detection is by measuring the edge asymmetry in the local density of states (LDoS). Within each of the bulk gaps, the LDoS will be strongly enhanced on only one side of a zigzag sample for almost all energies throughout that gap, as shown in Fig. \ref{fig: 3}e. This surface charge accumulation to one side of the sample is evidence that the zigzag termination generates a polarity in a macroscopic sample with open boundary conditions.  We note that there are no corresponding edge modes for the special case of an armchair configuration due to significant valley mixing, as shown in Fig. \ref{fig: 3}c.

 Having established the presence of edge modes on one side of a zigzag nanoribbon, we now explore different strain profiles which can give rise to the appropriate effective pseudomagnetic field for these edge modes to be observed \cite{VKG10}. As a first example, let us consider a strained flat sheet with $h=0.$ Periodic strain fields without vertical displacements must be accompanied by periodically-embedding regions of local compression ($\nabla \cdot \mathbf{u}< 0$) and extension ($\nabla \cdot \mathbf{u}> 0$). One example strain field of this type is $ u_x \propto   \sqrt{3}\cos \left( \mathbf{G}_2 \cdot \mathbf{r} \right)-\sqrt{3} \cos \left( \mathbf{G}_3 \cdot \mathbf{r} \right) $ and $u_y \propto   \cos \left( \mathbf{G}_2 \cdot \mathbf{r} \right)+ \cos \left( \mathbf{G}_3 \cdot \mathbf{r} \right) -2\cos \left( \mathbf{G}_1 \cdot \mathbf{r} \right).$ We notice that this strain field preserves $C_{3z}$ and breaks $C_{2z}$ as required. However, while flat strain fields contain all the necessary theoretical ingredients to produce the desired effect, they would not be easily achievable because of the large elastic-energy cost incurred as a result of the inevitable local compressive strain.

\begin{figure}
    \centering
    \includegraphics[width=3.4in]{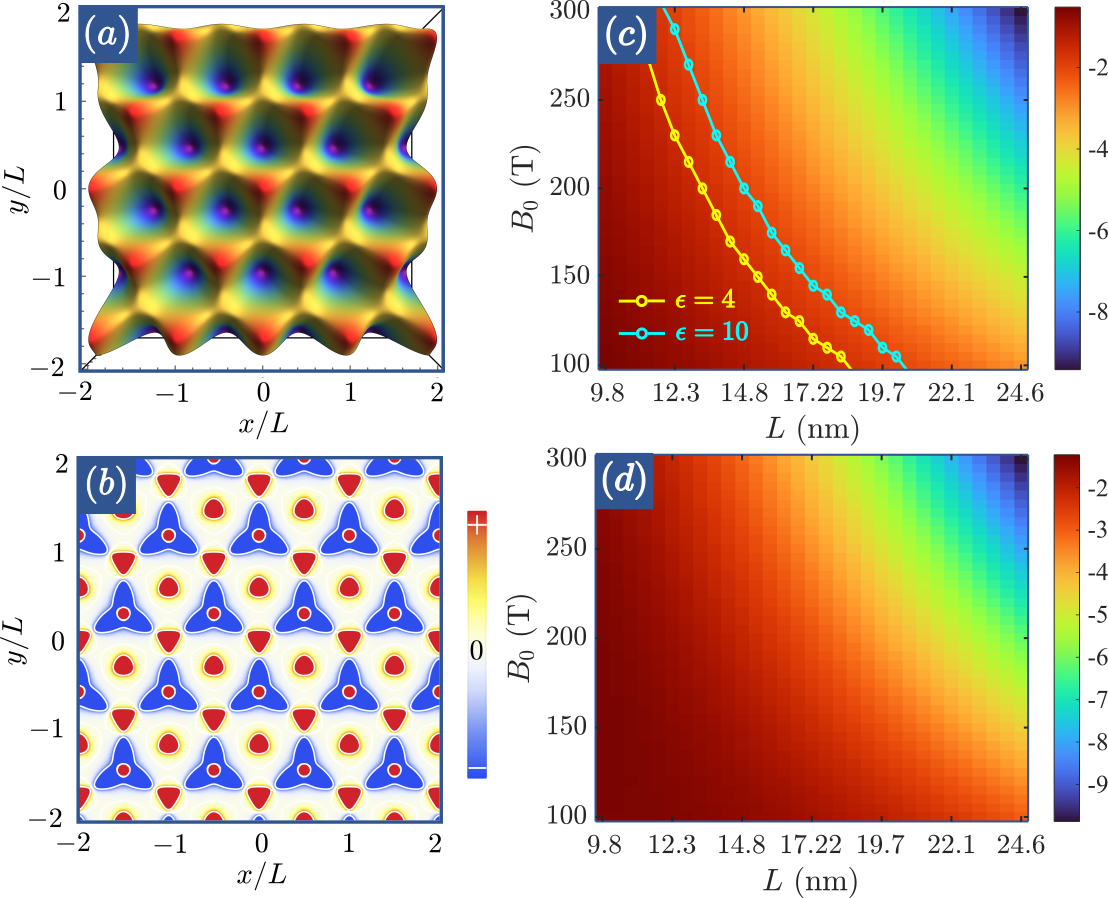}
    \caption{\textbf{Designing strain fields.} (a) $h(\mathbf{r})$ and (b) its Gaussian curvature with $\phi = -\pi/4$ as defined in the text. Following Ref. \cite{MA20}, we show (c) the estimated bandwidth of the energy manifold at charge neutrality and (d) the estimated magnitude of the first insulating gap above charge neutrality. The yellow and cyan lines show the Coulomb energy scale $E \sim 14.4  \text{ \AA eV}/\epsilon L$ for $\epsilon=4$ and $\epsilon=10$ respectively. The colormaps have units $\log_{10}(\text{eV}).$ }
    \label{fig: 4}
\end{figure}

We can alternatively obtain the desired pseudo-gauge field by lifting the model into the third dimension and imposing an appropriate height profile $h(\mathbf{r}).$ This approach offers more experimental control because the height profile can be  engineered simply by placing graphene on a substrate designed to produce a desired height profile $h(\mathbf{r})$. As the graphene membrane conforms to the substrate topography, it deforms not only vertically but laterally as well to minimize elastic energy. Taking into account relaxation in all three directions \cite{GHLD08,WBTKL08}, the following height profiles produce the desired pseudo-gauge fields, $h(\mathbf{r}) = h_0\sum_{i=1}^3 \cos \left( \mathbf{G}_i \cdot \mathbf{r} + \phi \right)$ with $\phi = \pm \pi/4.$ Roughly, this profile can be created in two ways: either by arranging \textit{triangular} pillars on a triangular lattice or by placing \textit{cylindrical}  pillars in a hexagonal lattice where the heights of the two sublattices are different, as shown in Fig. \ref{fig: 4}a. The plus $(+)$ and minus $(-)$ signs on $\phi$ indicate two complementary structures with opposite orientations, one features  boundary states on the top edge and the other on the bottom edge. These two structures cannot be locally transformed into one another.  This can be understood in terms of the Gaussian curvature. The local Gaussian curvature for $\phi = -\pi/4$ is shown in Fig. \ref{fig: 4}b. To get the complementary structure, we need to invert the local Gaussian curvature at various high-symmetry regions of the unit supercell. However, since the total curvature over one unit supercell is always zero due to the Gauss-Bonnet theorem, this can only be done by a global transformation of the structure that interchanges extrema and saddle points in the height profile. We note that since the Gaussian curvature is invariant under the reflection $h \mapsto -h,$ inverting the sign of the vertical deflection does not transform to a complementary structure. The pseudomagnetic field produced by this strain field has magnitude $B_0 \sim \left( 6 \times 10^5 \right) \times h_0^2/N^3$ T/\AA$^2.$ Using Fig. \ref{fig: 4}cd as a guide, $h_0$ can be chosen for a given $N$ to yield the desired bandwidths and gaps. For instance, for $N = 60$ and $B_0 = 100 $ T, we find $h_0 \approx 6$ \AA.

The preceding considerations show that in addition to the recently-realized platform using NbSe$_2,$ the edge-state physics we uncover here should be  accessible in many other graphene-based settings with appropriately-engineered substrate topographies. Namely, strain fields that break $C_{2z}$ can produce one-sided edge states. Actually, similar physics can be accessed using non-periodic strains as well, as described in Ref. \cite{LG10}. We emphasize that  while these edge states are fragile near avoided crossings in the presence of valley mixing,  we can exponentially suppress intervalley hybridization by increasing the superlattice period. Furthermore, disorder can of course lead to localization just as in the case of topological crystalline insulators where disorder generically breaks the spatial symmetries needed to protect edge states \cite{F11, AF15}. However, if disorder on average preserves valley symmetry, it is reasonable to expect these edge modes will persist as high-mobility transport channels \cite{AF15,NH19}. In this limit, we can still  regard valley as a good approximate quantum number. This strong valley-helical character of these edge modes is of intrinsic technological interest since it can potentially be harnessed for various valleytronic applications \cite{MP16,SYCR16,VN18}. Finally, it is worth considering in future works the possibility that the boundary physics explored here might be accessible in other two-dimensional materials as well.

We thank Antonio L. R. Manesco, Eva Andrei, Francisco Guinea, and Martin Claassen for useful discussions.  VTP acknowledges support from the NSF Graduate Research Fellowships Program and the P.D. Soros Fellowship for New Americans. EJM is supported by the Department of Energy under grant number DE-FG02-84ER45118.

\onecolumngrid
\setcounter{equation}{0}
\setcounter{figure}{0}
\renewcommand{\theequation}{S\arabic{equation}}
\renewcommand{\thefigure}{S\arabic{figure}}
\renewcommand{\bibnumfmt}[1]{[#1]}
\renewcommand{\citenumfont}[1]{#1}

\vspace{0.5in}

\begin{center}
\begin{Large}
Supplementary Material 
\end{Large}
\end{center}

\section{Graphene in a Periodic Magnetic Field}

\begin{figure}
    \centering
    \includegraphics[width=5in]{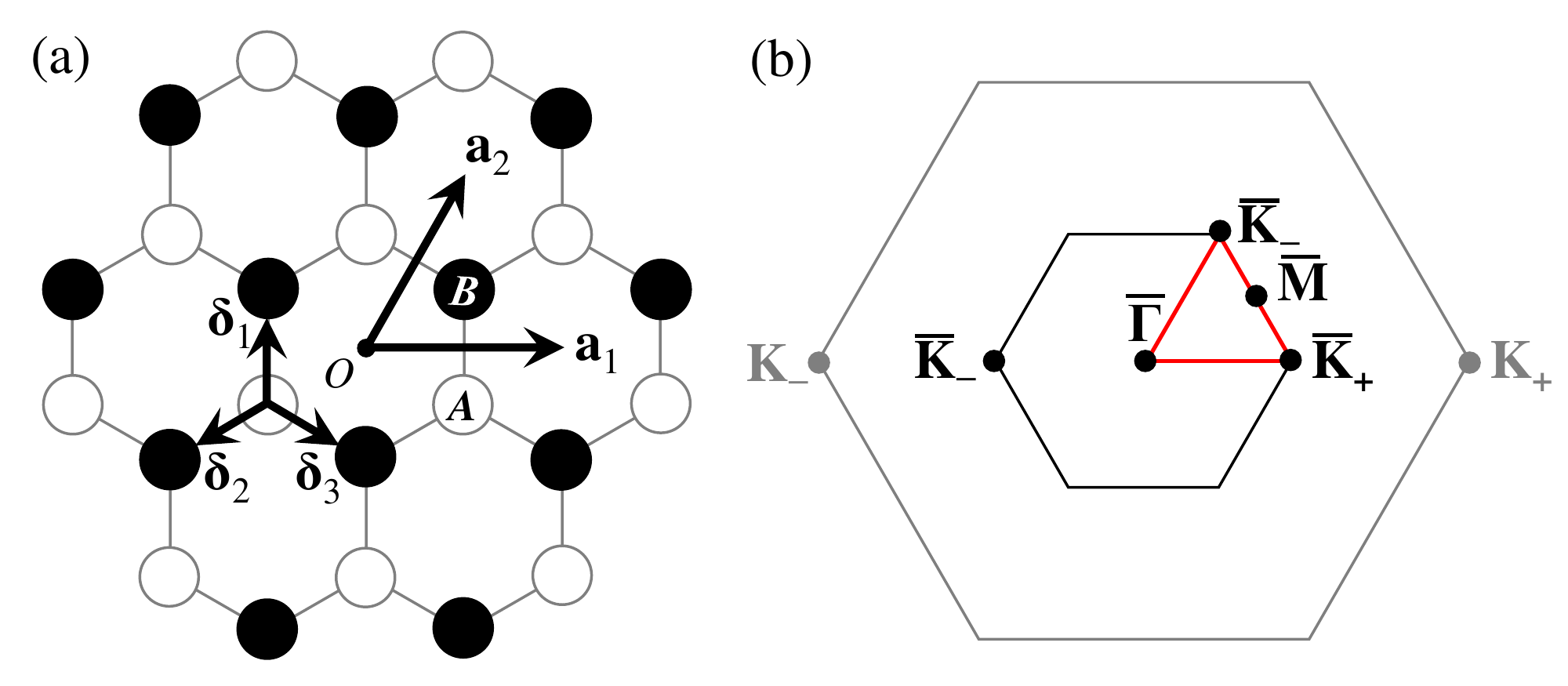}
    \caption{\textbf{Real-space and reciprocal-space representations of graphene lattice.} (b) The superlattice Brillouin zone is shown by the darker black lines, while the microscopic Brillouin zone is shown by the lighter gray lines.}
    \label{fig: lattice}
\end{figure}

We consider monolayer graphene in a \textit{real} periodic magnetic field. We denote the microscopic primitive lattice vectors as 
\begin{equation}
    \mathbf{a}_1 = a \left(1, 0 \right) \quad \text{and} \quad \mathbf{a}_2 = a \left( \frac{1}{2}  , \frac{\sqrt{3}}{2}\right),
\end{equation}
where $a = \sqrt{3}a_\text{CC}$ and $a_\text{CC}$ is the carbon-carbon distance. We choose the following microscopic reciprocal lattice vectors
\begin{equation}
    \mathbf{b}_1 = \frac{2\pi}{a} \left( 0, \frac{2}{\sqrt{3}} \right) \quad \text{and} \quad \mathbf{b}_2 = \frac{2 \pi}{a} \left(  -1, -\frac{1}{\sqrt{3}}\right).
\end{equation}
The inequivalent microscopic valleys are located at $\mathbf{K}_\pm= \left( \pm 4\pi/3a,0 \right).$ We assume for simplicity that the superlattice pattern generated by the magnetic field is commensurate with the microscopic structure. The superlattice primitive lattice vectors are given by $\mathbf{L}_i = N \mathbf{a}_i,$ where $N$ is an integer greater than unity. Likewise, the superlattice Brillouin zone (sBZ) is defined by $\mathbf{G}_i = \mathbf{b}_i/N.$ It is also useful to define the following nearest-neighbor vectors
\begin{equation}
    \boldsymbol{\delta}_1 = a_\text{CC} \left( 0,1 \right), \quad \boldsymbol{\delta}_2 = a_\text{CC} \left(- \frac{\sqrt{3}}{2}, -\frac{1}{2} \right), \quad \boldsymbol{\delta}_3 = a_\text{CC} \left(  \frac{\sqrt{3}}{2}, -\frac{1}{2} \right),
\end{equation}
as shown in Fig. \ref{fig: lattice}.

We take the real magnetic field to be 
\begin{equation}
    \mathbf{B}(\mathbf{r}) = B_0 \sum_{i=1}^3 \cos \left( \mathbf{G}_i \cdot \mathbf{r}  \right) \hat{z},
\end{equation}
where $\mathbf{G}_3 = -\mathbf{G}_1-\mathbf{G}_2.$ This magnetic field respects $C_{6z}$ rotation about the origin and preserves horizontal mirror symmetry, but it breaks all vertical mirror symmetries. A gauge for the magnetic vector potential $\mathbf{A}(\mathbf{r})$, $\mathbf{B} = \nabla \times\mathbf{A},$ which respects all of these symmetries is given by 
\begin{equation}
\begin{split}
        A_x(\mathbf{r}) &= - \frac{B_0 N a}{2 \pi} \frac{\sqrt{3}}{2} \left[ \sin \left(  \mathbf{G}_1 \cdot \mathbf{r}\right) - \frac{1}{2} \sin \left(\mathbf{G}_2 \cdot \mathbf{r} \right)- \frac{1}{2} \sin \left( \mathbf{G}_3 \cdot \mathbf{r}\right)\right], \\
        A_y(\mathbf{r}) &= - \frac{B_0 Na}{2 \pi} \frac{\sqrt{3}}{2} \left[ \frac{\sqrt{3}}{2} \sin \left( \mathbf{G}_2 \cdot \mathbf{r} \right) - \frac{\sqrt{3}}{2} \sin \left(\mathbf{G}_3 \cdot \mathbf{r}  \right) \right].
        \label{eq: magnetic vector potential}
\end{split}
\end{equation}
It is convenient to define the magnetic field strength in terms of a dimensionless variable $\delta t$ 
\begin{equation}
\label{eq: dimensionless constant}
    t_0\delta t = \frac{\sqrt{3} ev_F B_0 L}{4\pi},
\end{equation}
where $t_0 \approx 2.7$ eV is the nearest-neighbor hopping constant, $L=Na,$ and $\hbar v_F = \sqrt{3} t_0a/2$ \cite{CNG09}. Using Eq. \eqref{eq: dimensionless constant}, we can express the magnetic field in units of energy. We can write $\delta t /N = B_0/\bar{B},$ where $\bar{B}  = 8\pi \hbar/3a^2e \approx 90,000 $ T for $a = 2.46 $ \AA, the actual graphene lattice constant. In numerical calculation, we renormalize the lattice constant $\tilde{a} = \lambda a$ and $\tilde{t}_0 = t_0/\lambda$ to reduce computational cost while keeping $v_F$ fixed. For our purpose, we choose $\lambda =4.2.$ So $\tilde{B} = \bar{B}/\lambda^2 \approx 5,000$ T. In these renormalized units, we use $N \approx 14,$ $\delta t\approx 0.3,$ and $B_0 \approx 100$ T.

We implement the magnetic field in the tight-binding formalism using Peierls  substitution
\begin{equation}
    \hat{\mathcal{H}} = - \sum_{\alpha \beta} t_0\exp \left( -\frac{ei}{\hbar}  \int_{\Gamma \left[ \alpha \rightarrow \beta\right]} \mathbf{A}(\mathbf{r}) \cdot d\mathbf{r}\right) \hat{c}_{\beta}^\dagger \hat{c}_\alpha,
    \label{eq: tight-binding Hamiltonian supplementary}
\end{equation}
where $\Gamma\left[ \alpha \rightarrow \beta \right]$ is the straight path from $\mathbf{r}_\alpha$ to $\mathbf{r}_\beta.$ We use the convention that the electric charge is $-e$ with $e>0.$ The integral in Eq. \eqref{eq: tight-binding Hamiltonian supplementary} is simple enough to compute analytically for the nearest-neighbor hoppings
\begin{equation}
    -\frac{ei}{\hbar}\int^{\mathbf{R}_i}_{\mathbf{R}_i+ \boldsymbol{\delta}_j} \mathbf{A}(\mathbf{r}) \cdot d \mathbf{r} = \frac{i\sqrt{3} N\delta t }{2 \pi}\left\{\begin{array}{lr}
        \cos \left( \mathbf{G}_2 \cdot \mathbf{R}_i \right) - \cos \left( \mathbf{G}_3 \cdot \mathbf{R}_i \right)-\cos \left( \mathbf{G}_2 \cdot \mathbf{R}_i -\frac{2\pi}{3N}\right) +\cos \left( \mathbf{G}_3 \cdot \mathbf{R}_i -\frac{2\pi}{3N}\right), & \text{if } j=1\\
        \cos \left( \mathbf{G}_3 \cdot \mathbf{R}_i \right) - \cos \left( \mathbf{G}_1 \cdot \mathbf{R}_i \right)-\cos \left( \mathbf{G}_3 \cdot \mathbf{R}_i -\frac{2\pi}{3N}\right) +\cos \left( \mathbf{G}_1 \cdot \mathbf{R}_i -\frac{2\pi}{3N}\right), & \text{if } j=2\\
        \cos \left( \mathbf{G}_1 \cdot \mathbf{R}_i \right) - \cos \left( \mathbf{G}_2 \cdot \mathbf{R}_i \right)-\cos \left( \mathbf{G}_1 \cdot \mathbf{R}_i -\frac{2\pi}{3N}\right) +\cos \left( \mathbf{G}_2
        \cdot \mathbf{R}_i -\frac{2\pi}{3N}\right), & \text{if } j=3
        \end{array}\right. .
\end{equation}
As a check of consistency, in the $N \rightarrow \infty$ limit, we have $\cos \left( \mathbf{G}_k \cdot \mathbf{R}_i- \frac{2\pi}{3N} \right) = \cos \left( \mathbf{G}_k \cdot \mathbf{R}_i \right)\cos \left( \frac{2\pi}{3N} \right) +\sin \left( \mathbf{G}_k \cdot \mathbf{R}_i \right)\sin \left( \frac{2\pi}{3N} \right) \approx  \cos \left( \mathbf{G}_k \cdot \mathbf{R}_i \right)+\sin \left( \mathbf{G}_k \cdot \mathbf{R}_i \right)\frac{2\pi}{3N},$ and find 
\begin{equation}
     -\frac{ei}{\hbar}\int^{\mathbf{R}_i}_{\mathbf{R}_i+ \boldsymbol{\delta}_j} \mathbf{A}(\mathbf{r}) \cdot d \mathbf{r} \approx -\frac{i\delta t}{\sqrt{3}}\left\{\begin{array}{lr}
        \sin \left( \mathbf{G}_2 \cdot \mathbf{R}_i \right) - \sin \left( \mathbf{G}_3 \cdot \mathbf{R}_i \right), & \text{if } j=1\\
        \sin \left( \mathbf{G}_3 \cdot \mathbf{R}_i \right) - \sin \left( \mathbf{G}_1 \cdot \mathbf{R}_i \right), & \text{if } j=2\\
        \sin \left( \mathbf{G}_1 \cdot \mathbf{R}_i \right) - \sin \left( \mathbf{G}_2 \cdot \mathbf{R}_i \right), & \text{if } j=3
        \end{array}\right. = \frac{ei}{\hbar} \mathbf{A}(\mathbf{R}_i) \cdot \boldsymbol{\delta_j}.
\end{equation}
This approximation is true because $\mathbf{A}(\mathbf{r})$ does not vary significantly over the region of integration from $\mathbf{R}_i$ to $\mathbf{R}_i + \boldsymbol{\delta}_j$ so it can be pulled out of the integral sign.

\begin{figure}
    \centering
    \includegraphics[width=6.0in]{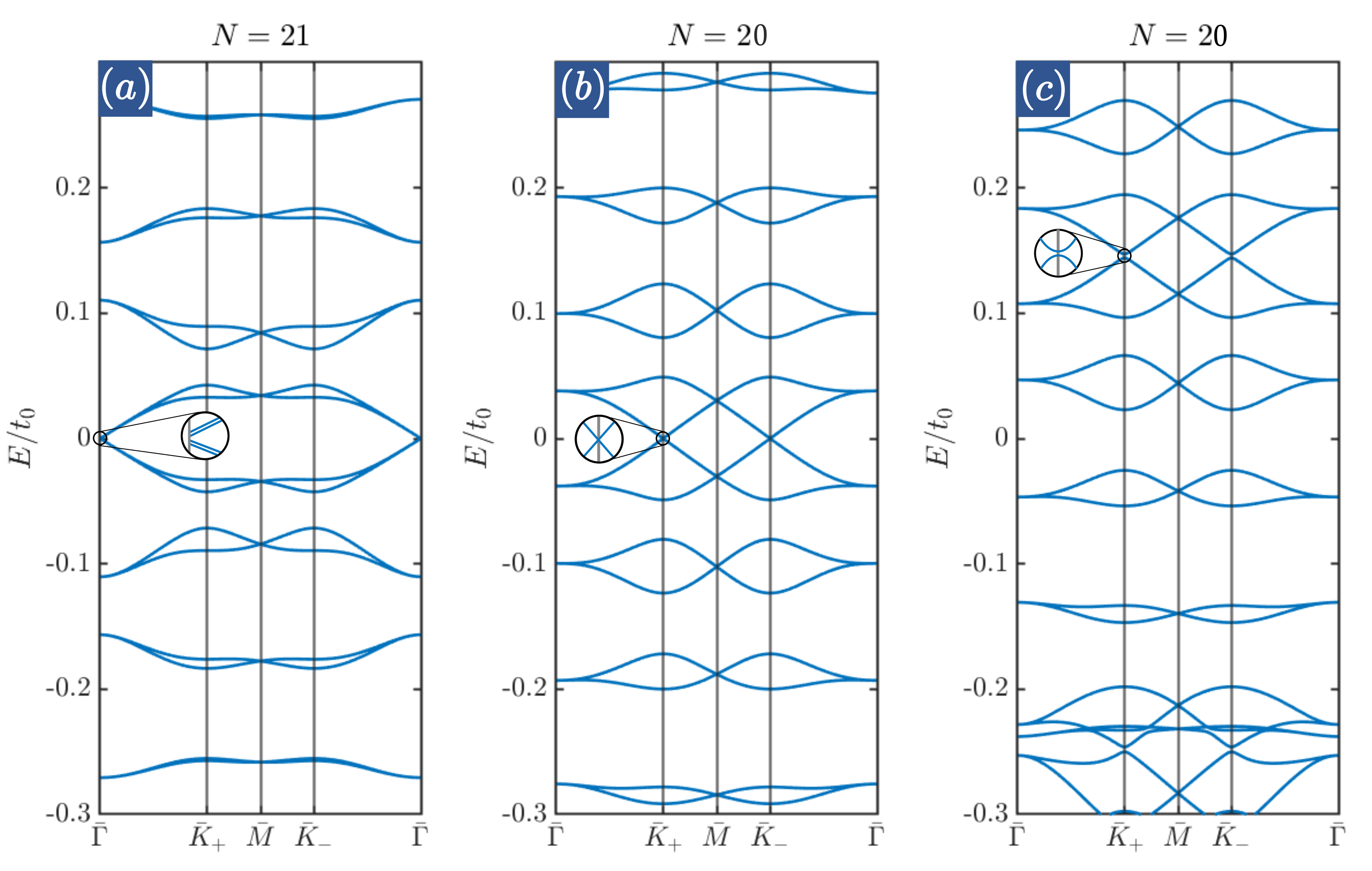}
    \caption{\textbf{Band structure of graphene in a real periodic magnetic field.} $\delta t = 0.2$ for all three plots. (a) Structure with $N = 21$ where the $\mathbf{K}_\pm$ Dirac cones are mapped to $\bar{\Gamma}$ and are gapped out by the superlattice potential. (b) Structure with $N = 20$ where the $\mathbf{K}_\pm$ Dirac cones are mapped to $\bar{K}_\pm$ and are \textit{not} gapped out by the superlattice potential. (c) Same structure as in (b) but with next-nearest-neighbor tunneling included. We see here that in addition to the absence of particle-hole symmetry, the otherwise gapless points at $\bar{K}_\pm$ also have avoided crossings.  }
    \label{fig:band structure}
\end{figure}

The presence of the magnetic field breaks all vertical mirror symmetries, but it preserves horizontal mirror symmetry and $C_{6z}$ symmetry about $\mathbf{r} = \mathbf{0}$. When this origin is aligned with a hexagon center of graphene, the whole system respects $C_{6z}$ symmetry. In the large-$N$ limit, we expect the low-energy physics will be insensitive to the precise microscopic alignment; so we can, without loss of generality, assume $C_{6z}$ symmetry will be effectively preserved. The gauge we have chosen for the magnetic vector potential preserves all of these symmetries. The Hamiltonian is invariant under the Abelian $C_{6h}$ group. Actually, it is invariant under a larger magnetic point group $C_{6h}\otimes \mathcal{T}M_x$. The periodic magnetic field defines a superlattice potential that folds the bands back onto themselves. In that process, where the bands intersect, strong hybridization can occur to gap out the degeneracies unless otherwise protected by a symmetry.  However, since the point symmetry group is the Abelian $C_{6h}$ group, there are no degeneracies protected by spatial point symmetries. For simplicity, we do not study further degeneracies that might occur at higher energies possibly protected by symmetries other than $C_{6h}$ since these are not relevant for our present purpose. We only worry about the degeneracies which seem to exist sometimes at $E = 0.$ To this end, we observe that there are  two classes of inequivalent structures with different behaviors at charge neutrality depending on the value of $N.$ Assuming that $\mathbf{K}_+$ either maps to $\bar{\Gamma}$ or $\bar{\mathbf{K}}_\pm,$ we find that 
\begin{equation}
\mathbf{K}_+ = n \left( \mathbf{G}_3-\mathbf{G}_2 \right) +  p\bar{\mathbf{K}}_+,
\end{equation}
where $n$ is an integer and $p = \lbrace -1,0,1 \rbrace.$ This implies that
\begin{equation}
    \frac{4\pi}{3a} =  \frac{4\pi n}{Na}+\frac{4\pi p}{3Na} \rightarrow N = 3n + p.
\end{equation}
If $N = 3n$ for some $n,$ then $\mathbf{K}_+$ is mapped to $\bar{\Gamma}.$ If $N = 3n\pm 1,$ then $\mathbf{K}_+$ is mapped to $\bar{\mathbf{K}}_\pm.$ Said another way, when $\mod \left(N,3 \right)=0,$ both $\mathbf{K}_+$ and $\mathbf{K}_-$ are mapped to the same $\bar{\Gamma}$ point. Thus, the  superlattice potential will generically gap out the Dirac cones at $\bar{\Gamma},$ as shown in Fig. \ref{fig:band structure}a. On the other hand, if $\mod \left(N,3 \right)=\pm 1,$ $\mathbf{K}_\pm$ are mapped to $\bar{K}_\pm,$ which are not related to each other by a reciprocal lattice vector. Thus, the Dirac cones descending from the original graphene band structure remain ungapped in this configuration, as shown in Fig. \ref{fig:band structure}b. However, it is worth nothing that these Dirac cones are not robust as in pristine graphene. Here, the existence of these Dirac cones is a consequence of the fictitious particle-hole symmetry in the nearest-neighbor approximation.  We demonstrate this explicitly by calculating the band structure with next-nearest-neighbor hoppings $t' \exp \left(-ei \int \mathbf{A} \cdot d\mathbf{r} /\hbar \right),$ where we take $t' = 0.05t_0.$ As before, we have 
\begin{equation}
\begin{split}
        &-\frac{ei}{\hbar}\int^{\mathbf{R}_i}_{\mathbf{R}_i+ \boldsymbol{\delta}_\text{nnn}} \mathbf{A}(\mathbf{r}) \cdot d \mathbf{r} \\
        &= \frac{i \delta t}{\sqrt{3}\pi}\left\{\begin{array}{lr}
         -2 \pi  \sin \left(\mathbf{G}_1 \cdot \mathbf{R}_i\right)+N \sin \left(\frac{\pi }{N}\right) \left(\sin \left(\mathbf{G}_2\cdot \mathbf{R}_i-\frac{\pi }{N} \right)+\sin \left(\mathbf{G}_3 \cdot \mathbf{R}_i+\frac{\pi }{N}\right)\right) ,  & \text{if } \boldsymbol{\delta}_\text{nnn} = \mathbf{a}_1 \\
        2 \pi  \sin \left(\mathbf{G}_1 \cdot \mathbf{R}_i\right)-N \sin \left(\frac{\pi }{N}\right) \left(\sin \left(\mathbf{G}_3 \cdot \mathbf{R}_i-\frac{\pi }{N}\right)+\sin \left(\mathbf{G}_2 \cdot \mathbf{R}_i+\frac{\pi }{N}\right)\right), & \text{if } \boldsymbol{\delta}_\text{nnn} = -\mathbf{a}_1 \\
        2 \pi  \sin \left(\mathbf{G}_3 \cdot \mathbf{R}_i\right)-N \sin \left(\frac{\pi }{N}\right) \left(\sin \left(\mathbf{G}_2 \cdot \mathbf{R}_i-\frac{\pi }{N}\right)+\sin \left(\mathbf{G}_1 \cdot \mathbf{R}_i+\frac{\pi }{N}\right)\right) ,& \text{if } \boldsymbol{\delta}_\text{nnn} = \mathbf{a}_2 \\
        -2 \pi  \sin \left(\mathbf{G}_3 \cdot \mathbf{R}_i\right)+N \sin \left(\frac{\pi }{N}\right) \left(\sin \left(\mathbf{G}_1 \cdot \mathbf{R}_i-\frac{\pi }{N}\right)+\sin \left(\mathbf{G}_2 \cdot \mathbf{R}_i+\frac{\pi }{N}\right)\right), & \text{if } \boldsymbol{\delta}_\text{nnn} = -\mathbf{a}_2 \\
        -2 \pi  \sin \left(\mathbf{G}_2 \cdot \mathbf{R}_i\right)+N \sin \left(\frac{\pi }{N}\right) \left(\sin \left(\mathbf{G}_3 \cdot \mathbf{R}_i-\frac{\pi }{N}\right)+\sin \left(\mathbf{G}_1 \cdot \mathbf{R}_i+\frac{\pi }{N}\right)\right), & \text{if } \boldsymbol{\delta}_\text{nnn} = \mathbf{a}_2-\mathbf{a}_1 \\
        2 \pi  \sin \left(\mathbf{G}_2 \cdot \mathbf{R}_i\right)-N \sin \left(\frac{\pi }{N}\right) \left(\sin \left(\mathbf{G}_1 \cdot \mathbf{R}_i-\frac{\pi }{N}\right)+\sin \left(\mathbf{G}_3 \cdot \mathbf{R}_i+\frac{\pi }{N}\right)\right) ,& \text{if } \boldsymbol{\delta}_\text{nnn} = \mathbf{a}_1-\mathbf{a}_2 \\
        \end{array}\right. .
\end{split}
\end{equation}
Since next-nearest-neighbor hoppings are always present in any realistic modeling because they are completely consistent with any space-time symmetry of the lattice, the Dirac cones which exist in the nearest-neighbor limit are not rigorously protected, as shown in Fig. \ref{fig:band structure}c. Nonetheless, for reasonable values of $t'$, which needs not be especially small, the nearest-neighbor limit yields a very good approximation to the low-energy physics \cite{CNG09}. As such, we will only keep nearest-neighbor hoppings henceforth. Besides the Dirac cones at charge neutrality, there are many small avoided crossings at higher energies. The small energy scale there is controlled by intervalley scattering, which is suppressed in the large-$N$ limit. Also, even though flat-band physics is not of our primary interest here, we remark in passing that as $|\delta t| \rightarrow 1,$ the low-energy bands become exceptionally flat.

\begin{figure}[h!]
    \centering
    \includegraphics[width=5.5in]{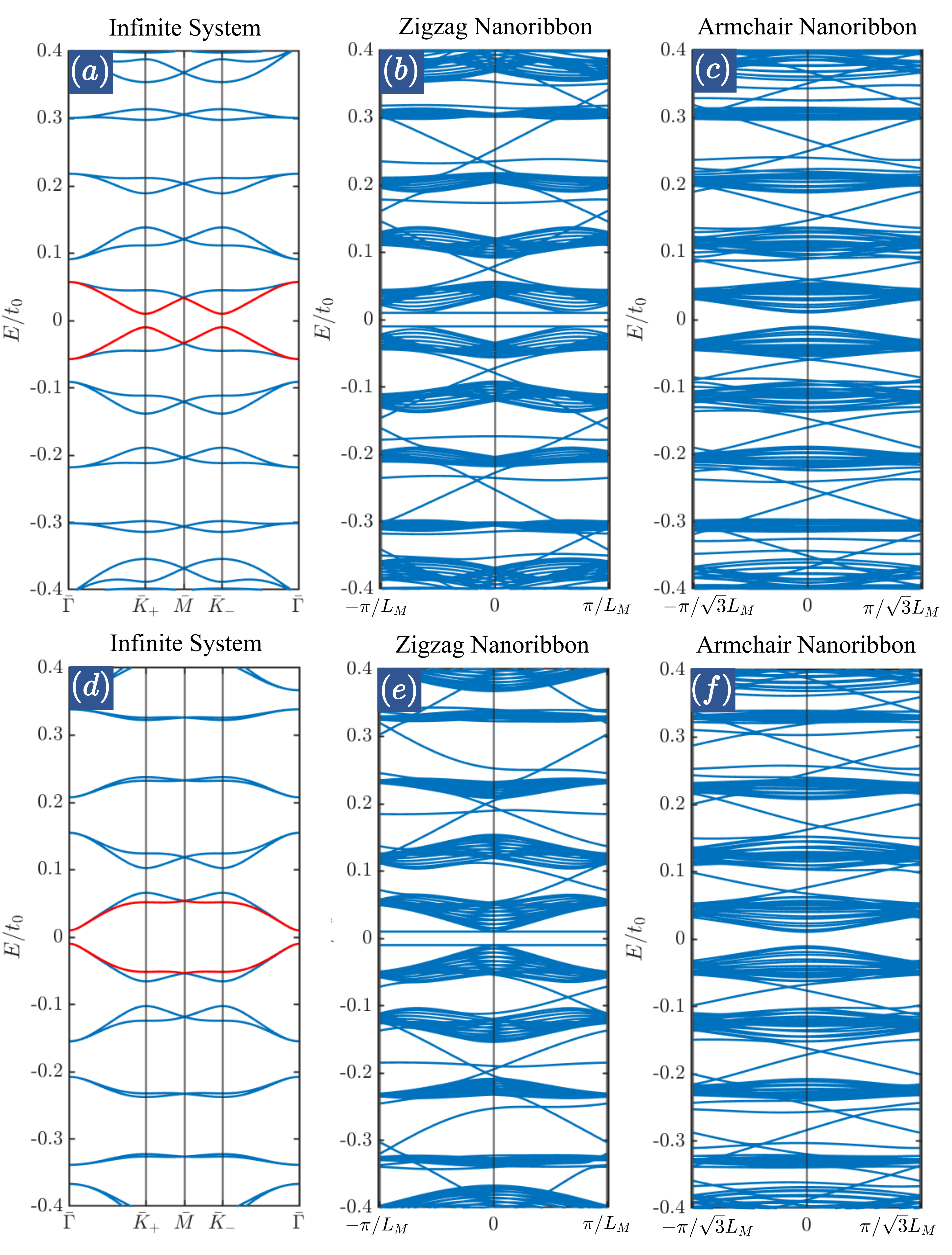}
    \caption{\textbf{Topology and edge states in graphene in a periodic magnetic field.} $N=19$ for (a)-(c), $N=18$ for (d)-(f), and $\delta t= 0.2$ for all plots. We add a small $\sigma_z$ term  of magnitude $0.01 t_0.$ In (a) and (d), band structures for infinite systems with red bands having Chern number $\mathcal{C} = 1$ and  blue bands having Chern number $\mathcal{C} = 0.$ Corresponding band structures for zigzag nanoribbons, (b) and (e), and for armchair nanoribbons, (c) and (f). We observe no edge states at $E=0,$ and a pair of counter-propagating edge states at the other insulating bulk gaps. }
    \label{fig:topology of bands}
\end{figure}

Next, we establish the topology of these bands using the Chern classification. We are only interested in the few insulating gaps above and below the charge neutrality point. To avoid the possible degeneracies at $E=0$, we add a small $\sigma_z$ sublattice potential that gaps out the Dirac cones at $\bar{K}_\pm.$ Because the sublattice potential is purely local, we know that this gap must be trivial and hosts no edge states for a sufficiently large $\sigma_z$. Its Chern number is zero \cite{H88}. Taken this as a reference, the Chern numbers associated with the insulating gaps away from zero can be calculated by using only bands from $E=0$ to $E = \mu,$ where $\mu$ is the chemical potential. This can then be confirmed by calculating the edge state spectrum. 

The Chern-number calculation for individual isolated bands is done numerically using a familiar formula \cite{F05,V18}
\begin{equation}
\label{eq: Chern number}
    \begin{split}
        \mathcal{C}_{\mathbf{k} = n \mathbf{g}_1+m\mathbf{g}_2} &= \Im \log \left(\bra{u_{\mathbf{k}}}\ket{u_{\mathbf{k}+\mathbf{g}_1}}\bra{u_{\mathbf{k}+\mathbf{g}_1}}\ket{u_{\mathbf{k}+\mathbf{g}_1+\mathbf{g}_2}}\bra{u_{\mathbf{k}+\mathbf{g}_1+\mathbf{g}_2}}\ket{u_{\mathbf{k}+\mathbf{g}_2}}\bra{u_{\mathbf{k}+\mathbf{g}_2}}\ket{u_{\mathbf{k}}}\right), \\
        \mathcal{C} &= \frac{1}{2\pi}\sum_{n,m=0}^{\mathcal{N}-1}\mathcal{C}_{n \mathbf{g}_1+m\mathbf{g}_2},
    \end{split}
\end{equation}
where $\ket{u_\mathbf{k}}$ is the periodic part of the Bloch wavefunction at $\mathbf{k},$ understood to belong to a band $b,$ and $\mathbf{g}_i=\mathbf{G}_i/\mathcal{N},$ where $\mathcal{N}^2$ is the size of the $\mathbf{k}$-space mesh. We impose $u_{\mathbf{k}+\mathbf{G}}=e^{-i\mathbf{G}\cdot \mathbf{r}} u_\mathbf{k}$ for any reciprocal lattice vector $\mathbf{G}.$ Because of the very small avoided crossings due to intervalley scattering, accurately computing the Chern number using Eq. \eqref{eq: Chern number} requires a fine grid. We use a $200 \times 200 = 4\times 10^4$ grid for all calculations of the Chern number. This is sufficient for the few bands in proximity to  $E=0,$ but may not be good enough for bands at higher energies since those are highly entangled. The results for a particular set of parameters are shown in Fig. \ref{fig:topology of bands}. We find that the two bands closest to charge neutrality have Chern number $\mathcal{C} = \pm 1.$ The few other bands above and below $E=0$ have $\mathcal{C} = 0.$ Therefore, the insulating gaps near charge neutrality all have $\mathcal{C} = \pm 1.$ The sign is selected by the sign of $B_0.$

\section{Continuum Theory with Strain Field}

In the limit that the superlattice period generated by the pseudomagnetic field is large compared to the microscopic lattice constant, the low-energy band structure is well-approximated by a continuum model where microscopic symmetries are neglected in favor of emergent superlattice-scale symmetries. In this regime, all the relevant physics is encapsulated by the Dirac cones at the $\mathbf{K}_+$ and $\mathbf{K}_-$ valleys. Assuming that the gauge field is spatially smooth, we neglect intervalley scattering entirely. Under these assumptions, the low-energy physics per valley is captured by a Dirac theory in the presence of a spatially-modulated gauge field. The unperturbed valley-projected Hamiltonian is 
\begin{equation}
\mathcal{H}_\nu^0(\mathbf{k}) = \hbar v_F \left( \mathbf{k} - \mathbf{K}_\nu \right) \cdot \left( \nu \sigma_x, \sigma_y \right),
\end{equation}
where $\mathbf{k}$ is measured from $\Gamma.$ In the presence of a gauge field, the Hamiltonian is modified using minimal coupling 
\begin{equation}
\mathbf{k} \mapsto \mathbf{k} + \frac{e}{\hbar} \mathbf{A}(\mathbf{r})
\end{equation}
to find 
\begin{equation}
\label{eq: S11}
\mathcal{H}_\nu(\mathbf{k}) = \mathcal{H}_\nu^0(\mathbf{k})  + ev_F \mathbf{A}(\mathbf{r) } \cdot \left( \nu \sigma_x, \sigma_y \right). 
\end{equation}
We note that using \eqref{eq: S11}, it can be demonstrated that the eigenstates are gauge-covariant and the energies are gauge-independent. Using a gauge transform $\mathbf{A}(\mathbf{r}) \mapsto \mathbf{A}(\mathbf{r}) - \nabla f(\mathbf{r}),$ the wavefunctions simply acquire a $U(1)$ phase $\psi(\mathbf{r}) \mapsto \exp \left( i ef(\mathbf{r})/ \hbar \right) \psi(\mathbf{r}).$ Therefore, at least in the continuum approximation, it matters not which gauge we choose to do the calculation. Here, it is implicit that $\bf{A}(\mathbf{r})$ is defined within one valley. The gauge field in the other valley is obtained by negation to recover $\mathcal{T}$ symmetry.

We now analyze the symmetries of the single-valley Hamiltonian \cite{SW19}. For this purpose, let us start with the point symmetry group of suspended pristine monolayer graphene, $D_{6h}.$ This group can be decomposed into a direct product $D_{6h} = C_{6v} \otimes M_z,$ where $M_z$ here is mirror symmetry $z \rightarrow -z$. In the presence of buckling due to a substrate, $M_z$ is broken; so the point symmetry group is lowered to at most $C_{6v}.$ This group is of order 12, and is generated by a sixfold rotation $C_{6z}$ and a reflection symmetry about the vertical plane $M_x$ that maps $x \mapsto -x.$ To determine which of these symmetries survive in the presence of a valley-projected pseudomagnetic field, we consider how these symmetries act on three degrees of freedom in the problem: sublattice $\sigma$, valley $\nu$, and spatial coordinate $\mathbf{r}.$ We observe the following properties of the generators of $C_{6v} \otimes \mathcal{T}:$
\begin{itemize}
\item $\mathcal{T}:$ exchanges valleys $\nu \mapsto -\nu,$ leaves sublattices invariant $\sigma \mapsto \sigma,$ leaves spatial coordinates invariant $\mathbf{r} \mapsto \mathbf{r},$ and flips the sign of the magnetic field $\mathbf{A} \mapsto - \mathbf{A}$.
\item $C_{6z}:$ exchanges valleys $\nu \mapsto -\nu,$ exchanges sublattices $\sigma \mapsto - \sigma,$ rotates spatial coordinates $\mathbf{r} \mapsto \mathcal{R}_{6z}\mathbf{r},$ and keeps the sign of the magnetic field $\mathbf{A} \mapsto \mathbf{A}.$  
\item $M_{x}:$ exchanges valleys $\nu \mapsto -\nu,$ keeps sublattices invariant $\sigma \mapsto  \sigma,$ reflects spatial coordinates $(x,y)\mapsto (-x,y),$ and flips the sign of the magnetic field $\mathbf{A} \mapsto -\mathbf{A}.$  
\end{itemize}
Using these properties, we find the following action of the generators in Table \ref{table I}.
\begin{table}
\begin{center}
  \begin{tabular}{ c | c | c | c}
    \hline\hline
     & $\nu \mapsto - \nu$ & $\sigma \mapsto - \sigma$ & $\mathbf{A} \mapsto - \mathbf{A}$ \\ \hline
    $C_{6z}$ & $\surd$ & $\surd$ & $\times$\\ \hline
    $C_{3z}$ & $\times$ & $\times$ & $\times$\\ \hline
    $M_{x}$ &$\surd$  &  $\times$ & $\surd$\\ \hline
    $\mathcal{T}$ & $\surd$ & $\times$& $\surd$\\ \hline
    \hline
  \end{tabular}
\end{center}
\caption{\textbf{Action of generators of $C_{6v} \otimes \mathcal{T}.$}}
\label{table I}
\end{table}

From this, we see that $C_{3z}$ is indeed a symmetry of the single-valley Hamiltonian since it keeps valley, sublattice, and pseudomagnetic field invariant. $M_y$ is the only other symmetry that does not flip valley, but it changes the sign of the gauge field; so it is not a symmetry of the single-valley Hamiltonian.  Of the other  valley-inverting symmetries, we find that $\mathcal{T} C_{2z}$ is violated because, although it preserves valley, it flips the sign of the pseudomagnetic field. On the other hand, $\mathcal{T}M_x$ keeps within a single valley and has the correct sign of gauge field. So it is a candidate for a good symmetry. From this heuristic preliminary analysis, we find that there are only two potential symmetries of the single-valley Hamiltonian: $C_{3z}$ and $\mathcal{T} M_x.$

We now show that these symmetries are preserved explicitly by constructing the corresponding symmetry operators on the single-valley Hilbert space.  To simplify notation, we take $\nu = +1$ valley for illustration; properties of the other valleys are obtained by applying $\mathcal{T}.$ The Hamiltonian in this valley in second-quantized language is 
\begin{equation}
\hat{\mathcal{H}} = \hbar v_F \int d^2 \mathbf{r} \hat{\psi}^\dagger(\mathbf{r}) \left[ -i \nabla_\mathbf{r}  + \frac{e}{\hbar} \mathbf{A}(\mathbf{r}) \right]\cdot \left( \sigma_x ,\sigma_y \right) \hat{\psi}(\mathbf{r}),
\end{equation}
where $\hat{\psi}(\mathbf{r}) = \left(\hat{\psi}_A(\mathbf{r}), \hat{\psi}_B(\mathbf{r})\right)^T $ is a two-component spinor. We now diagonalize this Hamiltonian to by writing  $\hat{\psi}(\mathbf{r})$ as a sum of plane-waves
\begin{equation}
\hat{\psi}(\mathbf{r})  = \frac{1}{(2 \pi)^2} \int_\text{sBZ} d^2\mathbf{k} e^{i \mathbf{k} \cdot \mathbf{r}} \sum_{\mathbf{G}} e^{i \mathbf{G} \cdot \mathbf{r}}  \hat{\psi}_\mathbf{k}(\mathbf{G}),
\end{equation}
where $\mathbf{k}$ is measured from $\bar{\Gamma},$ and sBZ stands for the superlattice  Brillouin zone. In our notation, the Dirac cone is located at  $\bar{\Gamma};$ however, this choice is arbitrary, and we are free to shift the Dirac cone around the superlattice Brillouin zone, especially when comparing to tight-binding calculations. We also have $\hat{\psi}_{\mathbf{k}+\mathbf{G}'}(\mathbf{G}) = \hat{\psi}_\mathbf{k}(\mathbf{G}+\mathbf{G}')$ for $\mathbf{k} \in \text{sBZ}.$  Substituting this into the Hamiltonian, we find 
\begin{equation}
\begin{split}
\hat{\mathcal{H}} &= \int \frac{d^2 \mathbf{k}}{(2 \pi)^2} \sum_{\mathbf{G}, \mathbf{G}'} \hat{\psi}^\dagger_{\mathbf{k}}(\mathbf{G}') \mathcal{H}_{\mathbf{G}',\mathbf{G}}(\mathbf{k}) \hat{\psi}_{\mathbf{k}}(\mathbf{G}), \\
\mathcal{H}_{\mathbf{G}',\mathbf{G}}(\mathbf{k}) &= \hbar v_F \delta_{\mathbf{G}',\mathbf{G}}\left( \mathbf{k} + \mathbf{G}\right)  \cdot \left( \sigma_x, \sigma_y \right)\\
&+ \left( a \delta_{\mathbf{G}',\mathbf{G}+\mathbf{G}_1 }+ a^\dagger \delta_{\mathbf{G}',\mathbf{G}-\mathbf{G}_1 } - \frac{a}{2}\delta_{ \mathbf{G}',\mathbf{G}+\mathbf{G}_2}- \frac{a^\dagger}{2}\delta_{\mathbf{G}',\mathbf{G}-\mathbf{G}_2 } -\frac{a}{2} \delta_{\mathbf{G}', \mathbf{G}-\mathbf{G}_1-\mathbf{G}_2}-\frac{a^\dagger}{2} \delta_{\mathbf{G}, \mathbf{G}+\mathbf{G}_1+\mathbf{G}_2} \right) \sigma_x \\
&+ \left( \frac{\sqrt{3}a}{2} \delta_{\mathbf{G}', \mathbf{G}+\mathbf{G}_2 }+ \frac{\sqrt{3}a^\dagger}{2} \delta_{\mathbf{G}',\mathbf{G}-\mathbf{G}_2}-\frac{\sqrt{3}a}{2} \delta_{ \mathbf{G}',\mathbf{G}-\mathbf{G}_1-\mathbf{G}_2}-\frac{\sqrt{3}a^\dagger}{2} \delta_{\mathbf{G}', \mathbf{G}+\mathbf{G}_1+\mathbf{G}_2} \right) \sigma_y,
\end{split}
\end{equation}
where $a = -t_0 \delta t/2 i$. In this representation, the first-quantized single-valley Hamiltonian is a formally-infinite matrix in the space of reciprocal lattice vectors. However, for physics near charge neutrality, we can choose only the vectors $|\mathbf{G}| \ll |\mathbf{K}_+|.$ This will capture all the essential physics near  chemical potential.

Now, we determine the symmetry operators on this Hilbert space.  First, we consider the anti-unitary operator $\mathcal{T} M_x$. In real space, we have
\begin{equation}
\left[ \hat{\mathcal{T}} \hat{M}_x \right] \hat{\psi}(\mathbf{r}) \left[ \hat{\mathcal{T}} \hat{M}_x \right] ^{-1} =  \hat{\psi}(M_x\mathbf{r}), \quad \quad \left[ \hat{\mathcal{T}} \hat{M}_x \right] \hat{\psi}_\mathbf{k} (\mathbf{G})\left[ \hat{\mathcal{T}} \hat{M}_x \right] ^{-1} = \hat{\psi}_{M_y \mathbf{k}} \left(M_y \mathbf{G} \right),  \quad \quad \hat{\mathcal{T}}i \hat{\mathcal{T}}^{-1} = -i.
\end{equation}
Invariance of the single-valley Hamiltonian follows from $\mathcal{H}^*_{M_y\mathbf{G}',M_y\mathbf{G}}(M_y\mathbf{k})  = \mathcal{H}_{\mathbf{G}, \mathbf{G}'}(\mathbf{k}).$  Next, we consider $C_{3z}$
\begin{equation}
\hat{C}_{3z}\hat{\psi}(\mathbf{r}) \hat{C}_{3z}^{-1} = e^{2 \pi i\sigma_z/3}\hat{\psi}\left(R^{-1}_{3z}\mathbf{r}\right), \quad \quad \hat{C}_{3z}\hat{\psi}_\mathbf{k}(\mathbf{G}) \hat{C}_{3z}^{-1} =  e^{2 \pi i\sigma_z/3} \hat{\psi}_{R_{3z}^{-1}\mathbf{k}}(R_{3z}^{-1}\mathbf{G}).
\end{equation}
where $R_{3z}$ is the rotation matrix that rotates $\mathbf{r}$ by  $2\pi/3.$ Invariance of the Hamiltnonian follows from $e^{-2\pi i \sigma_z /3}\mathcal{H}_{R_{3z}\mathbf{G}', R_{3z} \mathbf{G}} \left( R_{3z} \mathbf{k} \right) e^{2\pi i \sigma_z /3} = \mathcal{H}_{\mathbf{G}', \mathbf{G}}(\mathbf{k}).$ From these observations, it is clear that $C_{3z}$ is respected. Thus, we have shown that the single-valley Hamiltonians respect $C_{3z}$ and $\mathcal{T} M_x.$

We remark briefly that it is straightforward in the $\mathbf{k}$-space representation that $\mathcal{T} C_{2z}$ cannot be a good symmetry. The action of this symmetry is 
\begin{equation}
\left[\hat{\mathcal{T}}\hat{C}_{2z} \right] \hat{\psi}_\mathbf{k}(\mathbf{G}) \left[\hat{\mathcal{T}}\hat{C}_{2z}\right]^{-1} = \sigma_x \hat{\psi}_\mathbf{k}(\mathbf{G}),
\end{equation}
where $\sigma_x$ implements sublattice exchange. The following condition must be satisfied $\sigma_x \mathcal{H}^*_{\mathbf{G}' , \mathbf{G}}(\mathbf{k})\sigma_x = \mathcal{H}_{\mathbf{G}' , \mathbf{G}}(\mathbf{k})$ in order for this to be a symmetry. The combined effect of $\sigma_x$ and complex conjugation is to keep all the Pauli matrices invariant. However, complex conjugation also inverts the sign of the pseudomagnetic field. Hence, the Hamiltonian cannot be invariant under $\mathcal{T} C_{2z}$ as claimed earlier.

Before concluding this section, let us comment on an approximate symmetry that serves to ``protect" the Dirac points which descend from the perturbed Hamiltonian. We first justify why we should expect these Dirac points to be there without explicit calculation. For pristine monolayer graphene, as long as we are working with only two $p_z$ orbitals on a hexagonal lattice, the Dirac cones are guaranteed to be there by symmetries: a combination of parity and time-reversal symmetries eliminates the possibility of a $\sigma_z$ mass term, and threefold rotation pins these Dirac points to the zone corners. So, before  turning on the pseudomagnetic field, we take it as a given that the Dirac points are robust symmetry-protected crossings. When the perturbation is switched on, there is now a periodic potential that can hybridize bands which cross at the same energy if they are separated in momentum space by a reciprocal lattice vector. At zero energy and in the absence of intervalley scattering, there are no allowed scattering processes which mix the Dirac cones. At higher energies, there are states to hybridize and gap out accidental crossings. Indeed, this is generically what happens in the generation of flat bands. However, from a symmetry point-of-view, there is no longer $C_{2z}$ to protect these Dirac cones. So in what way are these Dirac cones  ``protected"? Indeed, they are not protected from the symmetries considered above. One can imagine perturbations which gap these Dirac points out without breaking any of the symmetries in $3 \underline{m}.$ Putting in a $\sigma_z$ term locally everywhere is one simple way of doing so. Another way is to include next-nearest neighbor hoppings. Yet, another way is to put on a perpendicular electric field that generates a periodic scalar potential since the graphene sheet is buckled.

With these considerations, we do not expect the Dirac cones to be protected. However, it is true practically that these Dirac cones are quite robust. It is, thus, desirable to consider a symmetry that prohibits a $\sigma_z$ term that generates a mass. Such a symmetry is particle-hole symmetry implemented in first-quantized form by $\sigma_z \mathcal{H} \sigma_z = - \mathcal{H}.$ This rules out any term proportional to $\sigma_z$ and the identity. So both a mass term and next-nearest neighbor hoppings are ruled out by this term. This symmetry is useful to protect the Dirac cones, but they are not essential to our work here since we are mostly interested in the higher energy band gaps.

\section{Tight-Binding Model with Strain Field}

We now construct a tight-binding representation of Eq. \eqref{eq: S11}. To do this, we need to include both valleys and recover $\mathcal{T}$ symmetry. In this case, the Hamiltonian must respect $\mathcal{T},$ 
$C_{3z},$ and $M_x$ symmetries. In the presence of a periodic strain field, the atoms are displaced slightly from their equilibrium positions. For sufficiently small displacements, the main effect to the band structure comes from the renormalization of the nearest-neighbor hopping integrals. We shall impose this assumption throughout, and continue to use the equilibrium positions of the orbitals, but adjust the hopping constants accordingly $t_{ij} \mapsto t_0 + \delta t_{ij}.$ The Hamiltonian is
\begin{equation}
\begin{split}
\mathcal{H} &= -\sum_{\langle ij \rangle} \left( t_0 + \delta t_{ij} \right) \hat{c}_i^\dagger \hat{c}_j ,
\end{split}
\end{equation}
where we have used $t_{ij} = t_{ji}$ because hopping between $p_z$ orbitals is a real process. Now, anticipating that at low energies, only states near  the microscopic $\mathbf{K}_+$ and $\mathbf{K}_-$ contribute to the physics, we project to this momentum sector to find that the Hamiltonian is altered by a vector potential of the form \cite{PCN09, CNG09, VKG10}, for the $\mathbf{K}_+$ valley,
\begin{equation}
\label{eq: vector potential}
A_x\left(\mathbf{r}_i\right) + i A_y\left(\mathbf{r}_i\right) \approx -\frac{1}{ev_F} \sum_{j} \delta t_j \left(\mathbf{r}_i \right) e^{-i \mathbf{K}_+ \cdot \boldsymbol{\delta}_j}.
\end{equation}
The vector potential at $\mathbf{K}_-$ is found by time reversing of the vector potential at $\mathbf{K}_+$ (by including an overall minus sign). In our convention, the valleys are located at $\mathbf{K}_\pm = \frac{4\pi}{3a} \left( \pm 1, 0 \right).$ If we coarse-grain the position so that the discrete locations $\mathbf{r}_i$ in the tight-binding basis can be replaced by a continuous displacement vector $\mathbf{r},$ we can define an effective magnetic field 
\begin{equation}
\mathbf{B}_\text{eff}(\mathbf{r}) = \nabla \times \mathbf{A}(\mathbf{r}).
\end{equation}
Note that if the strain field is known, then the corresponding vector potential is uniquely determined by Eq. \eqref{eq: vector potential}. However, if only the effective magnetic field is known, then the strain field admits a gauge ambiguity in the same way that a real magnetic potential also has gauge redundancy. In the latter case, we must be careful to study only gauge-independent properties at low energies as those are the only ones which can be reliably reproduced in practice.

\begin{figure}
\includegraphics[width=6in]{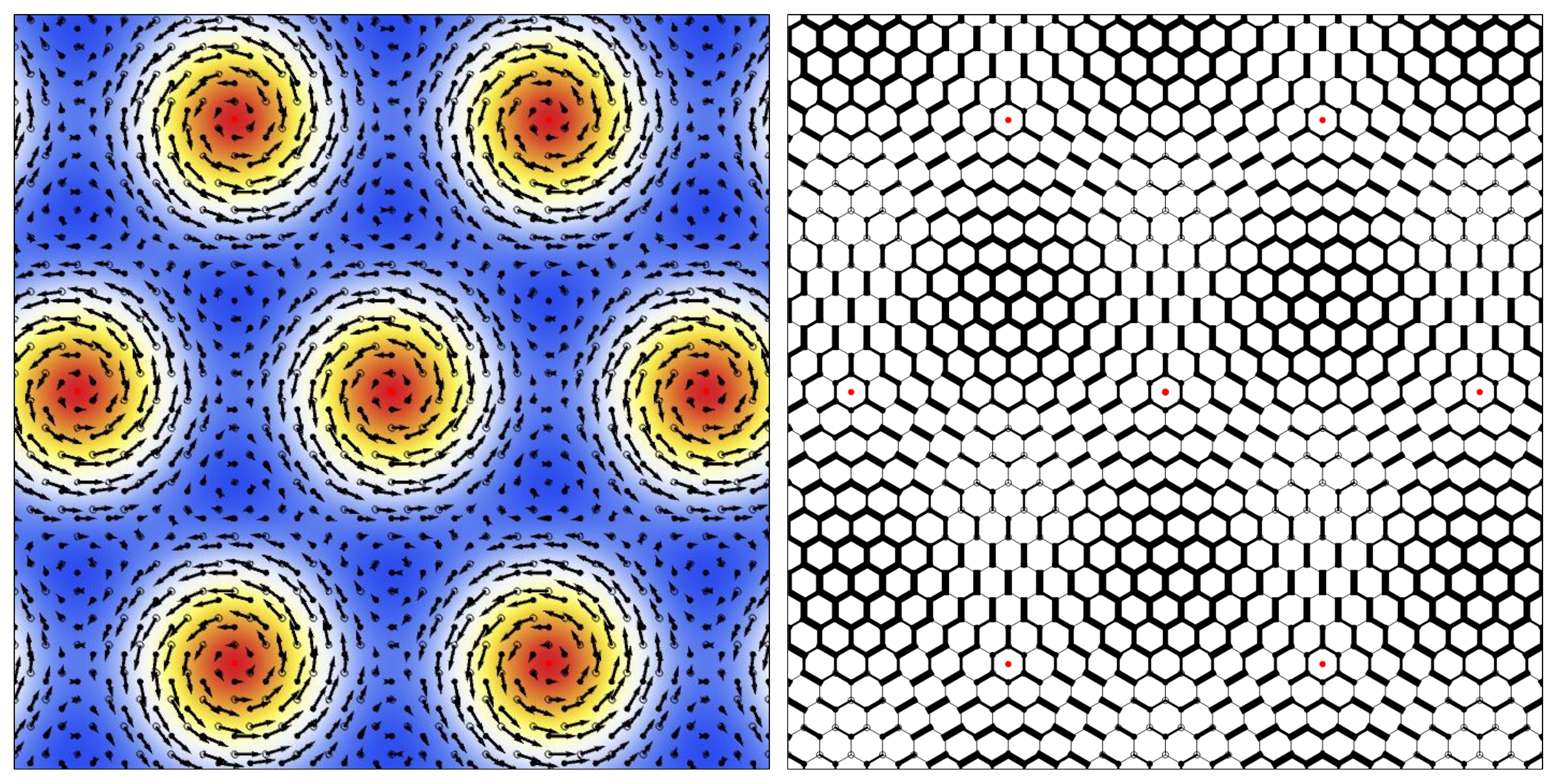}
\caption{\textbf{Symmetric representation of strain-induced pseudomagnetic field.} The left panel shows the intensity of magnetic field in color scale and a corresponding magnetic vector potential by the arrows. The right panel shows the $C_3$-symmetric realization of the corresponding strain field, where the strength of nearest-neighbor hoppings is indicated by the thickness of the lines.}
\label{fig: magnetic field}
\end{figure}

For our case of interest, the strain field is, in fact, not known. Only the effective magnetic field for a given valley is determinable experimentally \cite{MM20}.  Using the magnetic vector potential defined in Eq. \eqref{eq: magnetic vector potential}, Eq. \eqref{eq: vector potential} calculated on an $A$ site at $\mathbf{r}_i$ predicts that 
\begin{equation}
\label{eq: discrete potential}
\begin{split}
A_x(\mathbf{r}_i) &= -\frac{1}{ev_F} \left[ \delta t_1 (\mathbf{r}_i) - \frac{1}{2} \delta t_2(\mathbf{r}_i) - \frac{1}{2} \delta t_3(\mathbf{r}_i) \right] ,\\
A_y(\mathbf{r}_i) &= -\frac{1}{ev_F} \left[ \frac{\sqrt{3}}{2}\delta t_2(\mathbf{r}_i) - \frac{\sqrt{3}}{2} \delta t_3(\mathbf{r}_i) \right],
\end{split}
\end{equation}
where $\delta t_j (\mathbf{r}_i)$ is the change of bond strength in the $j^\text{th}$ direction. Comparing Eq. \eqref{eq: magnetic vector potential} with Eq. \eqref{eq: discrete potential}, one apparent choice for $\delta t_j (\mathbf{r}_i)$ is 
\begin{equation}
\label{eq: delta t}
\delta t_j \left( \mathbf{r}_i \right) =  \frac{ev_F B_0 L}{2 \pi} \frac{\sqrt{3}}{2} \sin \left( \mathbf{G}_j \cdot \mathbf{r}_i \right)  = t_0 \delta t \sin \left( \mathbf{G}_j \cdot \mathbf{r}_i \right).
\end{equation}
This gauge in Eq. \eqref{eq: delta t}  is similar to the one reported in Ref. \cite{ML21, ML20}, though we have chosen a different coordinate system. This choice of $\delta t_j(\mathbf{r}_i)$ respects $C_{3z}$ and $M_x$ symmetries but breaks $C_{2z}$ symmetry. This representation of the pseudomagnetic field and strain field is shown in Fig. \ref{fig: magnetic field}. Another convenient gauge, used in Ref. \cite{MA20}, is obtained by setting $A_y = 0$ and 
\begin{equation}
A_x \left(\mathbf{r} \right) = - \frac{B_0 L}{2 \pi} \left[ \frac{\sqrt{3}}{2} \sin \left(\mathbf{G}_1 \cdot \mathbf{r} \right) - \sqrt{3} \sin \left(\mathbf{G}_2 \cdot \mathbf{r} \right) - \sqrt{3} \sin \left(\mathbf{G}_3 \cdot \mathbf{r} \right) \right].
\end{equation}
Then, by Eq. \eqref{eq: discrete potential}, we must have $\delta t_2 \left( \mathbf{r}_i \right) = \delta t_3 \left( \mathbf{r}_i \right)$ and $-ev_F A_x (\mathbf{r}_i) = \delta t_1 (\mathbf{r}_i) - \delta t_2 (\mathbf{r}_i).$ From this, we can choose $\delta t_1(\mathbf{r}_i) = - \delta t_2 \left( \mathbf{r}_i \right)$ so that $
\delta t_1(\mathbf{r}_i) = -\frac{1}{2} ev_F A_x \left( \mathbf{r}_i \right).$ This gauge breaks $C_{3z}$ and $M_{x}$ so we will not use it in our work.

\begin{figure}
\includegraphics[width=7in]{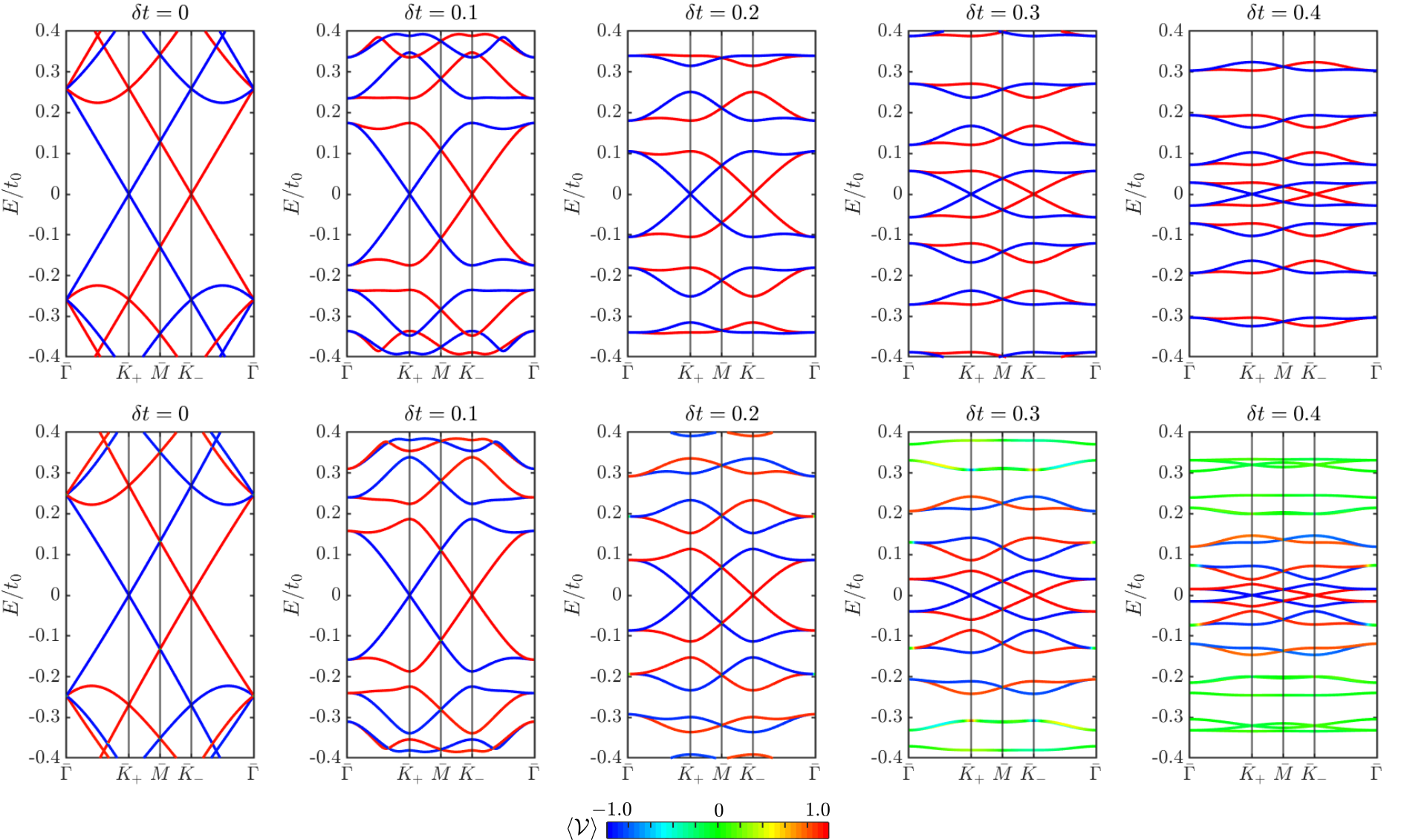}
\caption{\textbf{Comparison of continuum and tight-binding models.} The top plots are calculated using the continuum model, and the bottom plots are calculated using the tight-binding model. Here, $N = 14.$ We observe that the bands closest to $E=0$ are in qualitative, but not quantitative, agreement between the two models.}
\label{fig: comparison}
\end{figure}

\begin{figure}
\includegraphics[width=7in]{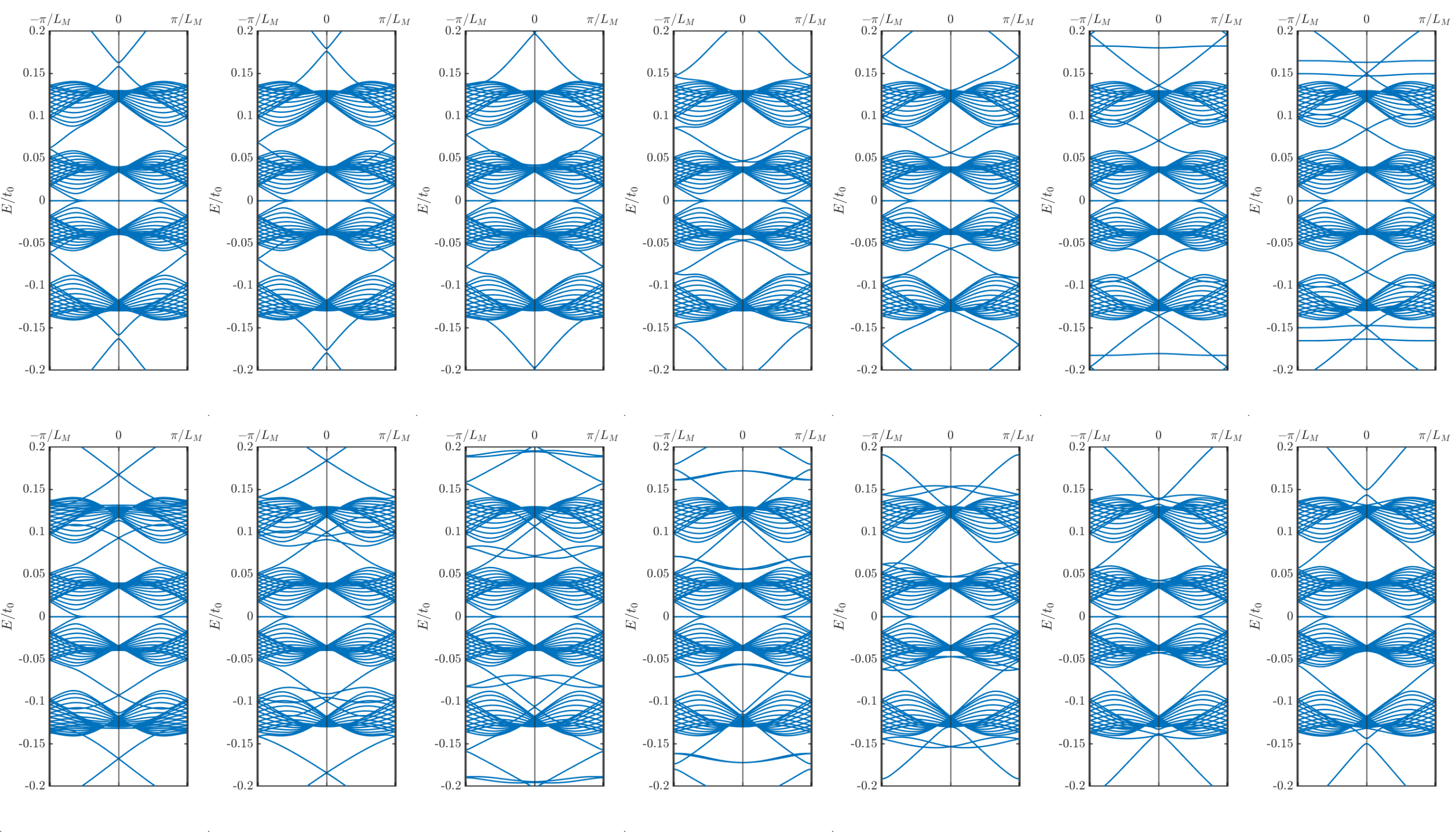}
\caption{\textbf{Robustness of edge states to lateral slip.} As we go from left to right and top to bottom, the zigzag nanoribbon of fixed length is moved along the $y$-direction by displacing the origin of the strain field by integer units of $\mathbf{a}_2$.  Here, $N=14;$ so after moving by $14\mathbf{a}_2,$ we return to the original structure. We observe that the one-sided edge states remain. }
\label{fig: lateral slip}
\end{figure}

Using Eq. \eqref{eq: discrete potential}, we can now calculate the strained-induced band structure. First, as a check of consistency, we compare the bands calculated from the continuum model and those calculated from the tight-binding model. Though we do not expect exact quantitative agreement between these two models, we expect qualitative agreement near  $E = 0$ in the large-$N$ limit. This is indeed the case, as illustrated in Fig. \ref{fig: comparison}. We can see from there that the continuum bands match qualitatively those of the tight-binding bands near $E=0.$ Namely, there is a set of doublet bands (per valley) at charge neutrality and singlet bands (per valley) elsewhere. For higher energies, these bands are no longer comparable, even qualitatively, because of increasingly stronger intervalley mixing and the emergence of non-Dirac dispersion. Reassured by this consistency check, we will use the tight-binding model from here onward to study edge modes on finite nanoribbons. In Fig. \ref{fig: lateral slip}, we simulate a zigzag nanoribbon with 150 carbon atoms across its width for $N = 14$ and $\delta t= 0.3$. Keeping the width fixed, we slide the the nanoribbon across the $y$-direction to assess the stability of the edge modes to precise termination. In other words, we move the origin of the strain field. We find that the existence of the one-sided edge modes is insensitive to this transformation. In the armchair configuration, there are no corresponding edge states because of strong valley hybridization.

\section{Designing Possible Strain Fields}

In this section, we consider two possible strain fields which give rise to the desired pseudomagnetic gauge field. Let the in-plane and out-of-plane displacement fields be denoted $(u_x(\mathbf{r}), u_y(\mathbf{r}))$ and $h(\mathbf{r})$ respectively. The strain field is defined by 
\begin{equation}
    u_{ij} = \frac{1}{2} \left( \partial_i u_j + \partial_j u_i + \partial_i h \partial_j h \right).
\end{equation}
The corresponding pseudomagnetic vector potential is \cite{SA02, M07, FMV13,MA20}
\begin{equation}
    \left( A_x, A_y \right) = -\frac{\hbar\beta}{2ea_\text{CC}} \left( u_{xx}-u_{yy}, -2u_{xy}  \right),
\end{equation}
where $\beta$ is a measure of how the bonds respond to being deformed. We use $\beta \approx 2-3.$ First, let us consider $h=0.$ It is straightforward to check that the following deformation field gives the appropriate pseudo-gauge field
\begin{equation}
\begin{split}
\label{eq: strain profile}
     \mathbf{u}(\mathbf{r}) &= -u_0 \left( \sqrt{3} \cos \left( \mathbf{G}_2\cdot \mathbf{r} \right) - \sqrt{3} \cos \left( \mathbf{G}_3 \cdot \mathbf{r}   \right), \cos \left( \mathbf{G}_2 \cdot \mathbf{r}   \right)+\cos \left( \mathbf{G}_3 \cdot \mathbf{r}   \right)-2\cos \left( \mathbf{G}_1 \cdot \mathbf{r}   \right)\right) ,  \\
     u_{xx}(\mathbf{r}) &= \frac{4 \sqrt{3} \pi  u_0 }{L}\cos \left(\frac{2 \pi  x}{L}\right) \sin \left(\frac{2 \pi  y}{\sqrt{3} L}\right) ,\\
     u_{yy}(\mathbf{r}) &= \frac{4 \pi  u_0 }{\sqrt{3} L}\sin \left(\frac{2 \pi  y}{\sqrt{3} L}\right) \left(\cos \left(\frac{2 \pi  x}{L}\right)-4 \cos \left(\frac{2 \pi  y}{\sqrt{3} L}\right)\right),\\
     u_{xy}(\mathbf{r}) &= \frac{4 \pi  u_0 }{L}\sin \left(\frac{2 \pi  x}{L}\right) \cos \left(\frac{2 \pi  y}{\sqrt{3} L}\right),\\
     B_0 & = \frac{16 \pi ^2 \beta \hbar  u_0}{3 ea_\text{CC} L^2} \approx \left(4\times 10^{5} \text{ T/\AA}\right) \times \frac{u_0 \beta}{N^2}.
\end{split}
\end{equation}
For rough estimates, we take $N\sim 40-100,$ $u_0 \sim 0.1 \times a_\text{CC} \approx 0.14$ \AA. This gives $B_0 \sim 10-100$ T. Although these estimates seem optimistic, the strain profile in Eq. \eqref{eq: strain profile} is practically difficult to achieve because it requires compression of the graphene sheet, which incurs high elastic energy costs. We can minimize this energy cost by lifting into the third dimension, allowing $h(\mathbf{r})$ to be non-zero. To do this, we assume that $h(\mathbf{r})$ is fixed by designing the appropriate substrate topography. Then, we find a corresponding $\mathbf{u}(\mathbf{r})$ for which the following elastic energy function is extremized \cite{GHLD08}
\begin{equation}
\label{eq: elastic energy}
    \mathcal{E}_\text{elas} = \int_{\Omega} d^2 \mathbf{r} \left[ \frac{\lambda}{2} \left( u_{xx}+u_{yy} \right)^2+ \mu \left( u_{xx}^2+u_{yy}^2+2u_{xy}^2  \right)\right],
\end{equation}
where $\Omega$ is the superlattice unit cell, and $\lambda$ and $\mu$ are elastic constants. Expanding Eq. \eqref{eq: elastic energy}, we obtain 
\begin{equation}
    \begin{split}
        \mathcal{E}_\text{elas} & = \int_\Omega d^2 \mathbf{r} \left[\left( \frac{\lambda}{2}+\mu \right) \left(u_{xx}^2+u_{yy}^2  \right) + \lambda u_{xx}u_{yy} +2\mu u_{xy}^2 \right] \\
        &= \int_\Omega d^2 \mathbf{r} \left[\left( \frac{\lambda}{2}+\mu \right) \left(\left( \frac{\partial u_x}{ \partial x} + \frac{f_{xx}}{2}  \right)^2+\left( \frac{\partial u_y}{ \partial y} + \frac{f_{yy}}{2}  \right)^2 \right) +  \lambda \left( \frac{\partial u_x}{ \partial x} + \frac{f_{xx}}{2}  \right)\left( \frac{\partial u_y}{ \partial y} + \frac{f_{yy}}{2}  \right)\right] \\
        &+ \int_\Omega d^2 \mathbf{r} \frac{\mu }{2} \left( \frac{\partial u_x}{\partial y} + \frac{\partial u_y}{ \partial x} + f_{xy} \right)^2,
    \end{split}
\end{equation}
where $f_{ij} = \partial_{i} h \partial_{j}h.$ Now, we expand $u_i$ and $f_{ij}$ using their Fourier series 
\begin{equation}
\begin{split}
    u_i(\mathbf{r}) &= \sum_{\mathbf{G}} \tilde{u}_i(\mathbf{G}) e^{i \mathbf{G} \cdot \mathbf{r}}, \quad \tilde{u}_i(\mathbf{G}) = \frac{1}{|\Omega|} \int d^2 \mathbf{r} u_i(\mathbf{r}) e^{-i \mathbf{G} \cdot \mathbf{r}} ,   \\
    f_{ij}(\mathbf{r}) &= \sum_{\mathbf{G}} \tilde{f}_{ij}(\mathbf{G}) e^{i \mathbf{G} \cdot \mathbf{r}}, \quad \tilde{f}_{ij}(\mathbf{G}) = \frac{1}{|\Omega|} \int d^2 \mathbf{r} f_{ij}(\mathbf{r}) e^{-i \mathbf{G} \cdot \mathbf{r}}  ,
\end{split}
\end{equation}
to obtain
\begin{equation}
\begin{split}
    \mathcal{E}_\text{elas} &= |\Omega|\sum_\mathbf{G}\left( \frac{\lambda}{2}+\mu \right) \left( iG_x \tilde{u}_x(\mathbf{G}) + \frac{\tilde{f}_{xx}(\mathbf{G})}{2}  \right)\left( -iG_x \tilde{u}_x^*(\mathbf{G}) + \frac{\tilde{f}_{xx}^*(\mathbf{G})}{2}  \right) \\
    &+|\Omega|\sum_\mathbf{G}\left( \frac{\lambda}{2}+\mu \right) \left( iG_y \tilde{u}_y(\mathbf{G}) + \frac{\tilde{f}_{yy}(\mathbf{G})}{2}  \right)\left( -iG_y \tilde{u}_y^*(\mathbf{G}) + \frac{\tilde{f}_{yy}^*(\mathbf{G})}{2}  \right) \\
    &+  |\Omega|\sum_\mathbf{G} \frac{\lambda}{2} \left[\left( iG_x \tilde{u}_x(\mathbf{G}) + \frac{\tilde{f}_{xx}(\mathbf{G})}{2}  \right)\left( -iG_y \tilde{u}_y^*(\mathbf{G}) + \frac{\tilde{f}_{yy}^*(\mathbf{G})}{2}  \right)+\left( -iG_x \tilde{u}_x^*(\mathbf{G}) + \frac{\tilde{f}_{xx}^*(\mathbf{G})}{2}  \right)\left( iG_y \tilde{u}_y(\mathbf{G}) + \frac{\tilde{f}_{yy}(\mathbf{G})}{2}  \right)\right] \\
    &+ |\Omega|\sum_\mathbf{G} \frac{\mu}{2}  \left( iG_y \tilde{u}_x (\mathbf{G}) + iG_x \tilde{u}_y(\mathbf{G}) + \tilde{f}_{xy}(\mathbf{G}) \right) \left( -iG_y \tilde{u}_x^* (\mathbf{G}) - iG_x \tilde{u}_y^*(\mathbf{G}) + \tilde{f}_{xy}^*(\mathbf{G}) \right),   
\end{split}
\end{equation}
where we have used that because $u_{i}(\mathbf{r})$ and $f_{ij}(\mathbf{r})$ are real, $\tilde{u}_i(\mathbf{G}) = \tilde{u}_i^*(-\mathbf{G})$ and $\tilde{f}_{ij}(\mathbf{G}) = \tilde{f}_{ij}^*(-\mathbf{G}).$ Now, we extremize $\mathcal{E}_\text{elas}$ with respect to $\tilde{u}_x^*(\mathbf{G})$ and $\tilde{u}_y^*(\mathbf{G})$ to find
\begin{equation}
    \begin{split}
        -iG_x\left(\frac{\lambda}{2}+ \mu \right) \left( i G_x \tilde{u}_x + \frac{\tilde{f}_{xx}}{2} \right) -i G_x \frac{\lambda}{2}\left( iG_y \tilde{u}_y + \frac{\tilde{f}_{yy}}{2}  \right)- i G_y\frac{\mu}{2}\left( iG_y \tilde{u}_x  + iG_x \tilde{u}_y + \tilde{f}_{xy} \right) & = 0, \\
        -iG_y\left( \frac{\lambda}{2} + \mu \right)  \left( iG_y \tilde{u}_y + \frac{\tilde{f}_{yy}}{2}  \right) - iG_y \frac{\lambda}{2}\left( iG_x \tilde{u}_x + \frac{\tilde{f}_{xx}}{2}  \right)- i G_x \frac{\mu }{2}\left( iG_y \tilde{u}_x  + iG_x \tilde{u}_y+ \tilde{f}_{xy} \right) &= 0.
    \end{split}
\end{equation}
Solving this, we obtain
\begin{equation}
    \tilde{u}_i = \frac{i }{2 G^4 (\lambda +2 \mu )}\left[\tilde{f}_{ii} G_i \left(G_i^2 (\lambda +2 \mu )+G_j^2 (3 \lambda +4 \mu )\right)+(\tilde{f}_{jj} G_i-2 \tilde{f}_{ij} G_j) \left(G_i^2 \lambda -G_j^2 (\lambda +2 \mu )\right)\right],
\end{equation}
where $G = \sqrt{G_x^2+G_y^2}$ and $\lbrace i,j \rbrace = \lbrace x,y \rbrace$ but $i \neq j.$ From this, we can find all the desired quantities. In particular, the associated pseudomagnetic field is given by \cite{GHLD08}
\begin{equation}
\begin{split}
    \mathbf{B} &= (\partial_x A_y - \partial_y A_x ) \hat{z}= -\frac{\hbar \beta}{2ea_\text{CC}} \left(-2 \frac{\partial u_{xy}}{\partial x} - \frac{\partial u_{xx}}{\partial y}+ \frac{\partial u_{yy}}{\partial y}  \right) \hat{z}   \\
    &\rightarrow -\frac{\hbar\beta}{2ea_\text{CC}}\frac{i G_y \left(3 G_x^2-G_y^2\right) (\lambda +\mu ) }{G^4 (\lambda +2 \mu )}\left( \tilde{f}_{xx} G_y^2-2 \tilde{f}_{xy} G_yG_x+\tilde{f}_{yy} G_x^2\right) \hat{z}.
\end{split}
\end{equation}
Using this, we study the strain induced by the following height profiles 
\begin{equation}
    h(\mathbf{r}) = h_0 \sum_{i = 1}^3 \cos \left( \mathbf{G}_i \cdot \mathbf{r} \pm \pi/4\right).
\end{equation}
The associated pseudomagnetic field is 
\begin{equation}
    \mathbf{B} = \pm  \frac{8 \pi ^3 \beta \hbar h_0^2 (\lambda +\mu ) }{\sqrt{3} e a_\text{CC} L^3 (\lambda +2 \mu )}\sum_{i=1}^{3}\cos \left(\mathbf{G}_i \cdot \mathbf{r}\right) \hat{z} = \pm B_0\sum_{i=1}^{3}\cos \left(\mathbf{G}_i \cdot \mathbf{r}\right) \hat{z}.
\end{equation}
Taking $\lambda \sim \mu \sim 1$ eV/\AA$^2,$ we estimate the magnitude of $B_0 \sim \left(6\times 10^5 \text{ T/\AA}^2 \right) \times h_0^2/N^3.$

\bibliography{strain}

\end{document}